\documentclass[twoside,english]{iopart}
\usepackage[T1]{fontenc}
\usepackage[latin9]{inputenc}
\usepackage{geometry}
\usepackage{amsbsy}
\usepackage{amstext}
\usepackage{graphicx}
\usepackage{esint}
\usepackage[numbers]{natbib}

\makeatletter
\usepackage{iopams}
\usepackage{setstack}

\newcommand{\eqref}[1]{(\ref{#1})}
\usepackage{iopams}

\makeatother

\usepackage{babel}
\begin{document}
\title{Current statistics in the q-boson zero range process}
\author{A.A. Trofimova$^{\ddagger,*,1}$ and A.M. Povolotsky$^{\dagger,*,2}$}
\address{{$^*$National Research University Higher School of Economics,
20 Myasnitskaya, 101000, Moscow, Russia}}
\address{{$\dagger$Bogoliubov Laboratory of Theoretical Physics, Joint Institute
for Nuclear Research, 141980, Dubna, Russia}}
\address{{$\ddagger$Center for Advanced Studies, Skoltech, 143026, Moscow, Russia}}
\eads{\mailto{$^1$nasta.trofimova@gmail.com}, \mailto{$^2$alexander.povolotsky@gmail.com}}

\begin{abstract}
We obtain exact formulas of the first two cumulants of particle current
in the q-boson zero range process on a ring via exact perturbative solution
of the TQ-equation. The result is represented as an infinite sum of
double contour integrals. We perform the asymptotic analysis of the
large system size limit $N\to\infty$ of the expressions obtained.
For $|q|\neq1$ the leading terms of the second cumulant reproduce
the $N^{3/2}$ scaling expected for models in the Kardar-Parisi-Zhang
universality class. The scaling $q\asymp\exp(-\alpha/\sqrt{N})\to1$
corresponds to the the crossover between the Kardar-Parisi-Zhang and
Edwards-Wilkinson universality classes. Under this scaling the sum
converges to an integral, resulting in the crossover scaling function
derived previously for the asymmetric simple exclusion process and
conjectured to be universal.
\end{abstract}
\maketitle

\section{Introduction}

Kardar-Parisi-Zhang (KPZ) and Edwards-Wilkinson (EW) universality
classes unify plenty natural phenomena like interface random growth
including the spread of fire \citep{Maunuksela}, \citep{Myllys},
the growth of bacterial colonies \citep{WIMM}, solidification, wetting
and instability fronts \citep{AMM,TS}. They are also believed to
capture universal features of the large scale behaviour of traffic
flows \citep{DSS} and polymers in random media \citep{Kardar}. For
a review see \citep{H-HZ} and references therein. Both classes were
extensively studied starting from the 1980s \citep{Krug}. The first
analytic efforts were concerned with corresponding stochastic PDEs
for the interface growth, the continuous models, which were expected
to give a universal description of the whole universality classes. 

Description of the universal behaviour of an interface governed by
the stochastic linear EW PDE \citep{EW} was relatively straightforward.
In particular, it was shown to be characterized by two independent
critical exponents, which can be chosen e.g. to be the roughness exponent
$\zeta$ and dynamical exponent $z$. In 1+1 dimension, we deal with
in this paper, they are $\zeta=1/2$ and $z=2$. The first one shows
that the stationary EW interface is like the Brownian motion. The
second suggests that the propagation of fluctuations is purely diffusive. 

The nonlinear KPZ equation resisted the exact analytic treatment for
other 25 years. Luckily, the exact critical exponents in $(1+1)$
dimensions were guessed from heuristic arguments already in the seminal
paper of Kardar, Parisi and Zhang \citep{KPZ}. The roughness exponent
$\zeta=1/2$ can be found from the fact that an addition of the non-linearity
that transforms the EW equation to the KPZ one does not affect its
stationary solution. However, off the stationary state the propagation
of fluctuations becomes highly nontrivial being characterized by $z=3/2$
exponent. The exponents $\zeta$ and $z$ can be translated to yet
another pair of exponents $\alpha=\zeta/z$ and $\beta=1/z,$ responsible
for fluctuation and correlation scales respectively. Correspondingly,
their values are $\alpha=1/4,\beta=1/2$ for EW and $\alpha=1/3,\beta=2/3$
for KPZ class. 

At that early stage the dimensional analysis, mode-coupling method
and the renormalization group applied to the KPZ equation were useful
for finding (at least heuristically) critical exponents and conjecturing
scaling hypotheses about KPZ universality \citep{Krug}. However,
the exact scaling functions were off the scope of those approaches.
This is when the integrable stochastic interacting particle systems
came on stage. Thinking about  interacting particles in 1+1 dimensions
we mean a stochastic diffusive or driven-diffusive particle system
subject to an uncorrelated random force with local inter-particle
interactions. The particle density field in such a system can be thought
of as a gradient of the associated interface height. Conversely, the
interface height is nothing but the time-integrated particle current
in the particle system. Using this correspondence, the results obtained
for a particle system can be translated to the interface language
and vice versa. In addition, the integrability implies a special mathematical
structure behind the stochastic process that allows one to obtain
exact results, which produce the universal functions in the scaling
limit.

The paradigmatic examples are the symmetric and asymmetric simple
exclusion processes abbreviated as SSEP and ASEP \citep{Liggett},
which together with associated ``solid-on-solid'' interfaces give
examples of systems belonging to the EW and KPZ universality classes
respectively. In these models particles perform symmetric or asymmetric
random walks on a 1D lattice subject to exclusion interaction, which
prevents two particles from coming to the same site. Results on these
models can roughly be divided into two groups. 

The first group consists of results about the stationary state and
the large time behaviour of finite systems like a periodic lattice or
a segment with particle reservoirs attached to the ends. Among them
are the stationary state density and current profiles \citep{DDM1992} and   correlation functions \cite{DE1993},
the large deviation functions of the particle density and current on a ring 
\citep{DerLeb,ADLW} and on a segment \cite{DLS,DLS2,GE,LM2011,LMV2012,LP2014}, e.t.c. In particular the current cumulants  were obtained, which we discuss below in detail.
 (For historical account  and references see reviews \citep{Derida review, Derrida rev 2,lazarescu}). 

The second group describes the transient dynamics in the infinite
system. These are the one-point current distributions for several
special types of initial conditions \citep{Johansson,Nagao,PS,BR}.
In the case of TASEP, the totally asymmetric version of ASEP, it was
also possible to find all the equal-time multipoint current distributions
\citep{Sasamoto,BFPS} as well as those for unequal times \citep{IS,PPS}
and space-time points on space-like paths \citep{BF,BFS,PPP}. Some progress
also was achieved for the time-like correlation functions \citep{Johansson two-time}. 

Recently, the finite time results were also extended to systems with
periodic boundary conditions \citep{P2015,P2015_1,P2016,BL1,BL2,BL3}, where one can study
the transition between transient and stationary regimes. The latter
set actually belongs to the intersection of the two mentioned groups. 

Though the mentioned results give a vast picture of the EW and KPZ
universality in 1+1 dimensions, we are still far from constructing
the general theory. This is why testing  the results obtained against
other interacting particle models with richer dynamics is of interest.
In this paper we address probably the next simplest interacting particle
model, the zero range process (ZRP) \citep{Liggett}. This is a particle
system on 1D lattice with totally asymmetric jumps, where the jump
probability depends only on the occupation number of the site of departure.
The peculiarity of this system, which reveals its simplicity already
at the level of the stationary state, is the factorized form of the
stationary measure in the infinite or periodic system \citep{Evans}.
Here we consider the continuous time q-boson ZRP, where the functional
form of the jump rates ensures the Bethe ansatz integrability of the
stochastic generator. The quantum integrable model based on the quantum
deformation of the boson algebra was first proposed by Bogoliubov
and Bullough \citep{BB}. It later was adapted by Sasamoto and Wadati
\citep{SW} to be considered as the stochastic interacting particle
model. One of the authors of this paper rediscovered it looking for
the example of ZRP solvable by the coordinate Bethe ansatz \citep{P2003}.
Surprisingly, it was found to be dual under particle-hole transformation
of the q-TASEP model that appeared much later as a degeneration of
the Macdonald process \citep{BC}.

The hopping rates of the model are parameterized by a real parameter
$q\in\mathbb{R}$. The model is believed to belong to the KPZ universality
class when $q\neq1$ and to the EW universality class when $q=1$.
Several results were obtained on the q-boson ZRP. The mean group velocity
and the diffusion coefficient for two particles on the infinite lattice
were calculated \citep{SW}. The scaling form of the large deviation
function of the particle current was obtained for the periodic lattice
in the large system size limit and $q\neq1$ \citep{P2003}. 

More recently several results on the non-stationary dynamics on the
infinite lattice were obtained. The spectral theory for the q-boson
ZRP on the infinite lattice was constructed in \citep{BCPS1}. In
particular the Green functions of the evolution operator \citep{KL2011}
and the exact expression of the leftmost particle position \citep{KL2014}
were obtained using the Bethe ansatz diagonalization of the Markov
generator \citep{P2003}. Also many interesting results were obtained
for q-TASEP, the dual partner of q-boson ZRP, using both the Markov
duality \citep{BCS2014} and the relation between q-TASEP and Macdonald
process \citep{BC}. In particular the q-Laplace transform of the
particle position was obtained \citep{BCF2014}, using which the scaling
limit of the distribution of particle position was shown to converge
to Tracy-Widom distribution \citep{BV2015}. Finally we mention that
several generalizations of the q-boson ZRP were proposed like discrete
time q-boson ZRP \citep{PM2006}, q-Hahn ZRP \citep{P2013,BCPS2015}
and q-PushASEP \citep{CP2015}. Also the multi-species version of
q-boson model was constructed \citep{T2015,KMMO2016}. The stationary
measure for the latter was obtained in the matrix product form and
used for determining of the current and density profile \citep{KM2017}. 

In the present paper we undertake a further study of the large time
regime of q-Boson ZRP on the ring of size $N$ evaluating the exact
diffusion coefficient for the particle position and the associated interface
height. The problem of calculation of the current cumulants in exclusion
processes has a long history. The diffusion coefficient was  first
obtained for TASEP on a ring \citep{DEM} and on a segment \citep{DEM2} using the matrix product ansatz \citep{DEHP}.
Later the matrix product technique was extended to ASEP \citep{DerMal}.
The whole large deviation function of the distance traveled by a particle
in TASEP, which in particular yielded all the exact scaled cumulants
including the diffusion coefficient, was derived using the Bethe ansatz
in \citep{DerLeb}. This solution used significantly a special structure
of the TASEP Bethe equations, which is not present in the more general
ASEP case. The large deviation function for the ASEP was constructed
in the large system size limit under a special KPZ scaling by the
method of asymptotic solution of the Bethe equations proposed in \citep{K,LK}.
The universal current cumulants in SSEP on the ring were obtained
asymptotically both from the Bethe ansatz and from the fluctuating
hydrodynamics in \citep{ADLW}. Technique based on asymptotic solutions
of the Bethe equations was also applied to evaluate the current large
deviation function in the ASEP on the segment with open ends \citep{GE}.
Finally, the approach to finding the exact expressions of current
cumulants based on the functional Bethe ansatz or T-Q Baxter equation
was developed for ASEP on the finite ring by Prolhac and Mallick in
\citep{PM2008}. The exact large deviation function for the ASEP on
a segment was also found by adapting the matrix product \citep{LM2011,LMV2012} and
using the T-Q Baxter equation \citep{LP2014,CN2018}. 

Here we apply the method developed by Prolhac and Mallick to derive the first two cumulants 
of the particle current in the q-boson  ZRP on the ring. 
Our interest is two-fold. First, it is the technical
aspect of the perturbative solution of T-Q equation and as a result
finding the exact form of the second current cumulant. Though both
the ASEP and q-boson ZRP are solved in terms of the similar trigonometric
Bethe ansatz, the concrete details are very different. While the
former is related to the two-dimensional representation of the quantum
affine algebra $U_{q}(\widehat{sl_{2}})$, the latter is constructed
in terms of its infinite-dimensional q-boson representation. These
facts reveal themselves in the structure of solution of the T-Q equation.
In both cases the solutions are given in terms of polynomial truncation
of the generating function of stationary weights. Complexity of this
function seems to depend crucially on the dimension of underlying
representation. In the ASEP case the single site weight is a simple
binomial and the weight of $N$ sites is its $N$-th power. In contrast,
for the q-boson ZRP the single site function is an infinite sum representing
the entire or meromorphic q-exponential function, of which the $N$-th
power is much harder to manipulate with. As a consequence the exact
expression of the diffusion coefficient obtained in \citep{DerMal}
is an explicit sum of quantities constructed of binomial coefficients.
In our case we were able to represent the final result in the form
of an infinite sum of double contour integrals, which being less explicit,
however, is well suited for asymptotic analysis. 

Second but not the least goal we pursue is to find the scaling limit of
the formulas obtained and, thus, to check the scaling hypothesis made
before on the basis of the analysis of EW and KPZ equations and ASEP.
As due to the universality the results of this type are meaningful  far beyond the specific exactly solvable model, we give here a brief account of the previous studies of the universal scaling behavior of cumulants of particle current and interface height.

In general  it is expected that in the infinite system of KPZ and
EW universality classes a particle moves subdiffusively, with fluctuations growing with
time $t$ as $t^{\alpha}.$ So does the height of associated interface.
However, in the finite periodic system of size $N$ at large time
$t\gg N^{z}$ particles move diffusively, i.e. the variance of the
fluctuations grows linearly with time with the proportionality factor,
named diffusion coefficient, that vanishes in the infinite system
size limit as $N^{2\zeta-z}$, i.e. $1/N$ for EW and as $1/\sqrt{N}$
for KPZ universality class (for details see review \cite{Krug} and discussion in the text). 
The latter power law was one of the first
demonstrations of KPZ scaling behaviour obtained from exact solution
\citep{DEM}. The universal power law form of   cumulants   of KPZ interface height of arbitrary order  was conjectured
 in  \citep{KMH,AF} basing on the  analysis of dimensions supplied with the scaling invariance arguments. Specifically 
 the model dependent dimensionful factors were predicted, which come with the 
power laws, while the universal dimensionless constants are to be
determined from exact solutions.  The program of determining the universal
constants  was first realized within the exact calculation of the second current cumulant  
 \citep{DEM,DerMal} and of the cumulants of arbitrary order \cite{DerLeb,LK} in TASEP and ASEP. 
 The asymptotic  scaling form of the cumulants was also verified at several other models \cite{P2003,PM2006, PPH2003, DPP2015}. 
The same program for the EW class was undertaken in the mentioned solution of SSEP in \cite{ADLW}
and extended to ASEP under $1/N$ weak asymmetry scaling \cite{PMal2009}. 

The pure power  laws for cumulants, which take place at the KPZ and EW fixed points,   are  expected to be connected 
by   universal scaling functions. One example of such a function was derived in \cite{DerMal}
from the exact formula for the diffusion coefficient in ASEP under weak  $1/\sqrt{N}$ asymmetry scaling.
Another candidate for the scaling function describing the crossover of the third cumulant was also derived under the $1/\sqrt{N}$ asymmetry scaling in \cite{Prol2008}.    

Note that  the crossover  between KPZ and EW universality classes is exactly the regime of applicability of the KPZ equation, the two extreme cases being attained in the infinity and zero  limits of  the non-linearity parameter or equivalently at the late and early stages of the time evolution respectively.    An alternative approach to the search for the universal cumulants and crossover scaling  functions for EW and KPZ classes exploits the relation of KPZ equation with the problem of polymer 
in random media \cite{Kardar}, which suggests that the corresponding interface height is distributed as the free energy of the latter.      
Treating the problem    within the framework of replica method reduces the calculation of $n$-replica polymer partition function to the quantum problem of $n$ Lieb-Liniger bosons with attractive delta-interaction \cite{LL}, solvable via the Bethe ansatz.  On the other hand, the expectation of the $n$-polymer free energy per unit length plays the role of the  rescaled cumulant generating function of the polymer free energy with   $n$ being  the parameter of the generating function. Then, derivation of the rescaled cumulants   proceeds via   the heuristic   trick  of the expansion of the result around $n=0$. The way of analytical continuation of the solution of   $n$-particle problem to the non-integer number of particles  was proposed by Brunet and Derrida in \cite{BD2000}. In particular, the second order of the small $n$ expansion of the $n$-polymer free energy reproduces the crossover function   obtained in \cite{PMal2009}. Also   the third order term has been obtained, a correspondence of which with the result of  \cite{Prol2008} for the third cumulant has not been studied yet. In the joint limit $n\to 0$ and boson interaction constant $c\to \infty$ considered in \cite{BD2000_1} the  $n$-polymer free energy attains the scaling form of the scaled cumulant generating function  of KPZ class from \cite{DerLeb}. 

To  complete the review of studies of the  universal cumulant behaviour we also  mention the later results on  the universal cumulants in a non-stationary setting. Specifically on the fluctuation scale the fluctuations of the  interface in the EW universality class  are purely Gaussian  \cite{EW} while those in the KPZ class are described by universal distributions dependent on the global shape of initial conditions, some of which were previously known from the theory of random matrices  \cite{Johansson,Nagao,PS,BR}. The crossover between KPZ and EW universality classes is described by the solution of the KPZ equation, which was initially obtained for the narrow wedge initial conditions both from the weakly asymmetric limit of ASEP \cite{SS2010,ACQ2011}   and from the polymer in random medium \cite{CDR2010}. Finally we mention the finite time results \citep{P2015,P2015_1,P2016,BL1,BL2,BL3} for TASEP in the ring geometry, which have the stationary fluctuations on the ring and the transient fluctuations in the infinite system as limiting cases.

In our paper we perform the asymptotic analysis of the  exact formula obtained for the second stationary current cumulant  in the q-boson ZRP both under in the  KPZ and crossover  scaling limits. The formula obtained in the KPZ regime     confirms the   earlier result \citep{P2003}. Under the crossover scaling we obtain the scaling
function interpolating between the EW and KPZ classes, which confirms
the universality of the expression conjectured from the ASEP solution
\citep{DerMal}.

Our paper is organized as follows. In section 2 we define the model,
sketch necessary information about its stationary state and current
cumulants and state the first main result of the paper, exact formula
of the diffusion coefficient. In section 3 we explain the mapping
of the particle system to the interface on the cylinder and survey
predictions based of the scaling hypotheses for such an interface
belonging to KPZ and EW classes. We conclude the section 3 formulating
the results of the asymptotic analysis, which confirms these conjectures.
In section 4 we outline the derivation of exact formulas of the two
first scaled cumulants of particle current. Section 5 is devoted to
asymptotic analysis. We summarize and conclude in section 6. The appendix
contains details of calculations, which were moved there for the purpose
of better readability of the main text.

\section{q-boson zero range process: model, stationary state and results}

\subsection{The model and its observables }

ZRP is a stochastic interacting particle system. We define it on a
periodic one dimensional lattice with $N$ sites (sites $i$ and $N+i$
are identical) and $p$ particles. Each lattice site can be occupied
by an integer number of particles $n_{i}\geq0$. A particle configuration
is specified by the set of occupation numbers $\boldsymbol{n}=\{n_{1},\dots n_{N}\}$.
The total number of configurations is $C_{N+p-1}^{p}$. 
\begin{figure}
\centering{}\includegraphics[width=0.8\linewidth]{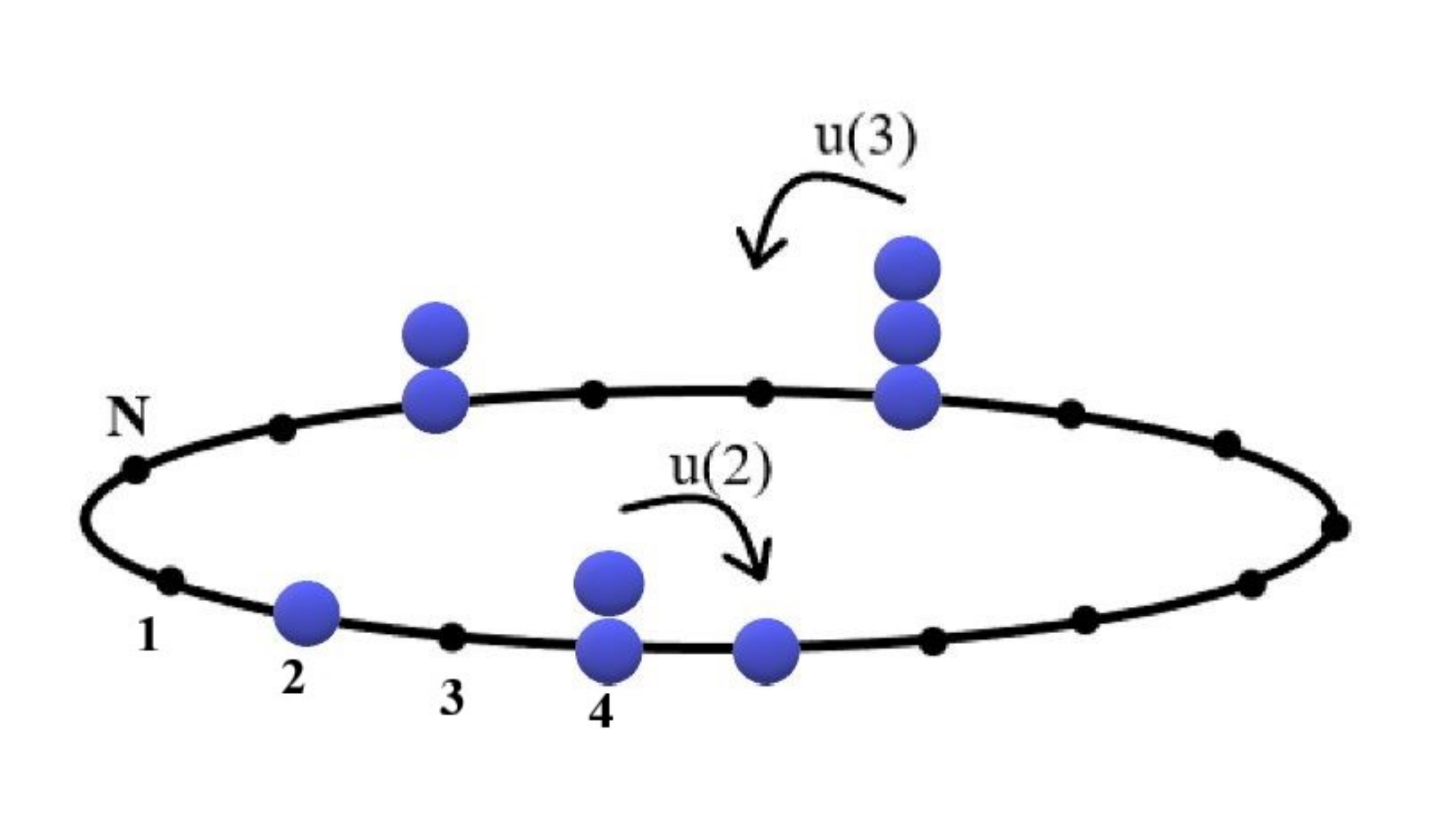}
\caption{$q$-boson zero range process on a ring with $N$ sites and $p=9$
particles.}
\label{ZRP_ring}
\end{figure}

We consider a continuous time Markov process on the set of particle
configurations. Each site has its own Poissonian alarm clock, which
rings with rate $u(n_{i})$. When the clock rings a particle from
site $i$ jumps to the neighbouring site $i+1$ (we imply that $u(0)=0$).

Let $P_{t}(C)$ be the probability for the system to be in configuration
$C$ at time $t$. The probability solves the master or forward Kolmogorov
equation 
\begin{equation}
\partial_{t}P_{t}(\boldsymbol{n})=\mathcal{L}P_{t}(\boldsymbol{n}).\label{ME}
\end{equation}
Here $\mathcal{L}$ is the operator, whose action on probability is
defined by 
\[
\mathcal{L}P_{t}(\boldsymbol{n})=\sum_{\boldsymbol{n}'}(u(\boldsymbol{n}'\rightarrow\boldsymbol{n})P_{t}(\boldsymbol{n}')-u(\boldsymbol{n}\rightarrow\boldsymbol{n}')P_{t}(\boldsymbol{n})),
\]
where the rate $u(\boldsymbol{n}'\rightarrow\boldsymbol{n})$ of transition
from configuration $\boldsymbol{n}'$ to $\boldsymbol{n}$ is equal
to $u(n'_{i})$ if the configuration $\boldsymbol{n}$ is obtained
from $\boldsymbol{n}'$ by a single jump of a particle from site $i$
the to the site $i+1$ and zero otherwise. In the following we will
deal with the particular choice of the rates 
\[
u(n)=[n]_{q}=\frac{1-q^{n}}{1-q},
\]
which was shown to be the one necessary for the Bethe ansatz integrability
\citep{P2003}. These rates are positive when $q>-1.$ This is the
range we consider below.

Having a solution of the master equation corresponding to particular
initial conditions one can compute the expectation of any function
of configuration at given time. Often one would also like to study
the statistics of additive functionals on trajectories of the process,
by which we mean a quantity $Y_{t}$ changing its value by a fixed
amount $\delta Y_{\boldsymbol{n}'\rightarrow\boldsymbol{n}}$ every
time the system jumps from $\boldsymbol{n}'$ to $\boldsymbol{n}$.
To this end, one considers the joint probability $P_{t}(\boldsymbol{n},Y)$
for the configuration to be $\boldsymbol{n}$ and the value of $Y_{t}$
to be $Y$ at time $t$. Its generating function 
\[
G_{t}(\boldsymbol{n},\gamma)=\sum_{Y=0}^{\infty} P_t(\boldsymbol{n},Y)e^{\gamma Y}
\]
is a solution of the non-stochastic deformation of (\ref{ME}) 
\[
\partial_{t}G_{t}(\boldsymbol{n},\gamma)=\mathcal{L_{\gamma}}G_{t}(\boldsymbol{n},\gamma),
\]
where the matrix of the deformed operator $\mathcal{L}_{\gamma}$
is obtained from that of $\mathcal{L}$ by multiplying every off-diagonal
element corresponding to transition from $\boldsymbol{n}'$ to $\boldsymbol{n}$
by $e^{\gamma\delta Y_{\boldsymbol{n}'\rightarrow\boldsymbol{n}}}.$
We consider a particular example of $Y_{t}$, the total distance traveled
by all particles by time $t.$ In this case the increase of $Y_{t}$
due to jump of a single particle is always $\delta Y_{\boldsymbol{n}'\rightarrow\boldsymbol{n}}=1$,
so that the action of $\mathcal{L}_{\gamma}$ is as follows.
\begin{equation}
\mathcal{L_{\gamma}}G_{t}(\boldsymbol{n,\gamma})=\sum_{\boldsymbol{n}'}(e^{\gamma}u(\boldsymbol{n}'\rightarrow\boldsymbol{n},\gamma)G_{t}(\boldsymbol{n}',\gamma)-u(\boldsymbol{n}\rightarrow\boldsymbol{n}')G_{t}(\boldsymbol{n},\gamma))\label{ME_for_F}
\end{equation}
The moment generating function of the random variable $Y_{t}$ is
given in terms of $G_{t}(\boldsymbol{n},\gamma)$.
\[
\mathbb{E}e^{\gamma Y_{t}}=\sum_{\boldsymbol{n}}G_{t}(\boldsymbol{n},\gamma)
\]
The utility of $G_{t}(\boldsymbol{n},\gamma)$ reveals itself in an
observation that in the long time limit its behaviour is dominated
by the largest eigenvalue $\lambda(\gamma)$ of matrix $\mathcal{L_{\gamma}},$
\[
\lambda(\gamma)=\lim_{t\rightarrow\infty}\frac{\ln\mathbb{E}e^{\gamma Y_{t}}}{t}=\sum_{n=1}^{\infty}c_{n}\frac{\gamma^{n}}{n!},
\]
i.e the function $\lambda(\gamma)$ plays the role of the generating
function of scaled cumulants 
\[
c_{n}=\lim_{t\to\infty}\frac{\big\langle Y_{t}^{n}\big\rangle_{c}}{t}
\]
of $Y_{t}$, where we use notation $\big\langle\xi^{n}\big\rangle_{c}$
for $n-$th cumulant of the random variable $\xi$. In particular,
the first two scaled cumulants, which we deal with below, have a simple
physical meaning. The first one
\begin{eqnarray*}
J=J(N,p) & : & =c_{1}=\lambda'(0)
\end{eqnarray*}
is the expected number of particle jumps in the system per unit time,
aka mean integral particle current, obtained by time-averaging of
the expected total number of jumps made by all particles by time $t$
growing to infinity. The second scaled cumulant is the group diffusion
coefficient 
\[
\Delta=\Delta(N,p):=c_{2}=\lambda''(0)
\]
associated with the joint motion of all particles. Instead of these
\textit{extensive} quantities associated with the whole system we
can consider \textit{intensive} quantities associated with a single
bond or a particle. In a sense the intensive quantities are more convenient
to characterize the infinite system. 

For example, one can consider the number of particles $y_{i}(t)$
that have passed through the bond $(i,i+1)$ by time $t$. Obviously
the variation of the number of particles passing through different
bonds on the ring is bounded by the total number of particles, i.e.
$y_{i}(t)=Y_{t}/N+O(N)$, where $O(N)$ is time independent. Hence,
the first two scaled cumulants of $y_{i}(t)$, the mean current and
current diffusion coefficient, are independent of the bond being equal
to 
\[
j_{N}(\rho=p/N):=\lim_{t\to\infty}\frac{\big\langle y_{i}(t)\big\rangle_{c}}{t}=\frac{J}{N}.
\]
\[
\Delta_{N}^{\mathrm{j}}(\rho):=\lim_{t\to\infty}\frac{\big\langle y_{i}^{2}(t)\big\rangle_{c}}{t}=\frac{\Delta}{N^{2}}.
\]
Suppose also that particles on the ring always preserve their order.
For example we may think that the particles in every site are arranged
into a column. Only the top particle is allowed to jump, while the
particle which jumps into a site takes the lowest position. Then,
the same argument applies to the coordinate of a particle $x_{i}(t).$
Therefore, the first two scaled cumulants of $x_{i}(t),$ the mean
particle velocity and the single particle diffusion coefficient, are
\[
\mathrm{v^{p}}_{N}(\rho)=j_{N}(\rho)/\rho,\quad\Delta_{N}^{\mathrm{p}}(\rho)=\Delta_{N}^{\mathrm{j}}(\rho)/\rho^{2}.
\]
As will be seen below the first intensive cumulants approach finite
nonzero values in the thermodynamic limit, while the diffusion coefficients
vanish indicating that the corresponding quantities evolve subdiffusively
in the infinite system.

\subsection{Stationary state and scaled current cumulants\label{subsec:Stationary-state-and}}

Here, before stating the results we recap briefly the properties of
the stationary state of the q-boson ZRP and fix necessary notations.

The peculiarity of ZRP is the factorized form of the stationary probability
distribution \citep{Evans}, which makes the analysis of the stationary
state particularly simple. This is to say that the probability of
finding the system in a configuration $\boldsymbol{n}$ is given by
a product of one-site weights
\begin{equation}
\boldsymbol{P}_{st}(\boldsymbol{n})=\frac{\prod_{i=1}^{N}f(n_{i})}{Z(N,p)},\label{P_st}
\end{equation}
where the one-site weight is given by 
\begin{equation}
f(m)=\left\{ \begin{array}{cc}
\prod_{j=1}^{m}\frac{1}{u(j)}, & m>0\\
1, & m=0
\end{array}\right.,
\end{equation}
and 
\begin{equation}
Z(N,p)=\sum_{\left\{ \boldsymbol{n}:n_{1}+\cdots+n_{N}=p\right\} }\prod_{i=1}^{N}f(n_{i}).\label{def_Z}
\end{equation}
is the normalization factor referred to as the (canonical) partition
function. The partition function can be given an integral representation
with the use of the generating function of one-site weights 
\[
F(z)=\sum_{n=0}^{\infty}f(n)z^{n}.
\]
In the case of q-boson ZRP the series $F(z)$ is convergent for $z$
in the disk $\left|z\right|<1/(1-q)$ when $|q|\leq1$ and for $z$
in the whole complex plain when $|q|>1$ to infinite products, which
give two different q-exponential functions. 
\begin{equation}
F(z)=\left\{ \begin{array}{ll}
e_{q}(z):=(z(1-q);q)_{\infty}^{-1}, & |q|<1,\\
E_{1/q}(z):=(z(q^{-1}-1);q^{-1})_{\infty}, & |q|>1,
\end{array}\right.\label{qexp}
\end{equation}
where $(z;q)_{\infty}=\prod_{i=0}^{\infty}(1-zq^{i}).$ From the statistical
physics point of view the generating function of weights of particle
configurations in $N-$site system, $F^{N}(z)$, considered as a function
of fugacity $z$ can be thought of as the grand-canonical partition
function. The canonical partition function $Z(N,p)$ has a contour
integral representation
\begin{equation}
Z(N,p)=\oint\frac{F^{N}(z)}{z^{p+1}}\frac{dz}{2\pi i},\label{eq: def_Z_sum}
\end{equation}
which is a standard relation between the canonical and grand-canonical
partition functions. 

The partition function $Z(N,p)$ is a basic object in our consideration,
which will frequently appear further. The generating function approach
to its calculation is also useful to evaluation of various stationary
state observables. For example the cumulants of the occupation number
in the stationary state are given by 
\[
\left\langle n_{i}^{k}\right\rangle _{c}=\left.\frac{d^{k}}{d\gamma^{k}}\ln\oint\frac{F^{N-1}(z)F(e^{\gamma}z)}{z^{p+1}}\frac{dz}{2\pi i}\right|_{\gamma=0},
\]
while for the mean current through the bond $(i,i+1)$, which is the
mean number of particles jumping out of the site $i$, we have 
\begin{equation}
j_{N}=\mathbb{E}u(n_{i})=\frac{Z(N,p-1)}{Z(N,p)}.\label{def j}
\end{equation}
These quantities manifestly do not depend on the site $i$. Thus for
the integrated current we have 
\begin{equation}
J=\mathbb{E}\left(\sum_{i=1}^{N}u(n_{i})\right)=N\frac{Z(N,p-1)}{Z(N,p)}.\label{eq:J}
\end{equation}

The second scaled cumulant, aka diffusion coefficient, can not be
obtained from the simple stationary state analysis, being the simplest
observable that implicitly contains unequal time correlations. To
find this quantity we need to address first the full dynamical problem.
This is what will be done below. Here we give the final expression,
which is the main result of this article.

\textbf{Result 1. }The group diffusion coefficient $\Delta$ has the
following representation 
\begin{eqnarray}
\Delta & = & pJ+\frac{2N^{2}}{Z(N,p)^{2}}\oint\frac{dy}{2\pi i}\frac{F^{N}(y)}{y^{p}}\oint\limits _{|y|<|t|}\frac{dt}{2\pi i}\frac{F^{N}(t)}{t^{p}}\frac{\phi(y)}{t-y}\nonumber \\
 & + & \frac{2N^{2}}{Z(N,p)^{2}}\sum_{i=1}^{\infty}\oint\frac{dt}{2\pi i}\frac{F^{N}(t)}{t^{p}}\oint\limits _{|y|<|t|}\frac{dy}{2\pi i}\frac{F^{N}(y)}{y^{p}}\frac{q^{\pm i}\phi(yq^{\pm i})+\phi(y)}{t-yq^{\pm i}},\label{Answer}
\end{eqnarray}
 where plus and minus signs in the powers of $q$ correspond to $\left|q\right|<1$
and $q>1$ respectively and the integration contours are two nested
simple counterclockwise loops around the origin, which do not contain
any other poles. Also we defined the function
\begin{equation}
\phi(z)=\frac{J}{p}(\ln F(z))'-1.\label{def_phi}
\end{equation}
where we use notation $(\ln F(z))' = \partial_z (\ln F(z))$ for derivative of function $\ln F(z)$.

The formula (\ref{Answer}) is valid for $-1<q\neq1$. Taking the
limit $q\to1$ is a nontrivial exercise. It is, however, straightforward
to show that at $q=1$
\begin{equation}
J(N,p)=\Delta(N,p)=p.\label{eq:q=00003D1 curren and diff coef}
\end{equation}
Indeed, in this case a particle jumps from a site with the rate $u(n)=n.$
If we forget about the order of particles, we may think that every
particles tries to jump with unit rate independently of the others.
Then the above result trivially follows from the properties of the
sum of $p$ independent Poisson processes. We do not consider taking
the limit of the exact formula here. Rather we will demonstrate that
it is restored from the scaling limit below. 

For $q\neq1$ it is not difficult to use these formulas to see that
for $p=1$ particle on the lattice 
\[
J(N,1)=1,\quad\Delta(N,1)=1,
\]
which, as expected, do not depend on the lattice size $N$. A little
more work is needed to find these quantities in the two-particle case
\begin{eqnarray*}
J(N,2) & = & \frac{2N}{N+\left(1-q\right)2_{q}^{-1}},\\
\Delta(N,2) & = & \frac{4N}{3(N+\left(1-q\right)2_{q}^{-1})^{3}}\frac{3Nq+(2N+1)(N+1)3_{q}}{(1+q)^{2}},
\end{eqnarray*}
of which the limit $N\to\infty$ yields
\[
J(\infty,2)=2,\quad\Delta(\infty,2)=2+\frac{2}{3}\frac{(1-q)^{2}}{\left(1+q\right)^{2}}.
\]
where $n_{q}=\frac{1-q^{n}}{1-q}.$ The $q=0$ limit $\Delta=8/3$
agrees with the result obtained for TASEP \citep{DerLeb}, while the
$q\to1$ limit $\Delta=2$ corresponds to the non-interacting particle
picture discussed above. In principle it is possible to analyze also
the lager values of $p$, though the complexity quickly increases. 

Of course of physical interest is the behaviour of the cumulants in
the thermodynamic limit, in which the notion of universality becomes
relevant. Below we study the asymptotic limit of the exact formulas
to demonstrate that they fit with existing conjectures on the universality
in KPZ and EW universality class.

\section{Interface growth and KPZ-EW universality}

\begin{figure}
\centering{}\includegraphics[width=0.7\linewidth]{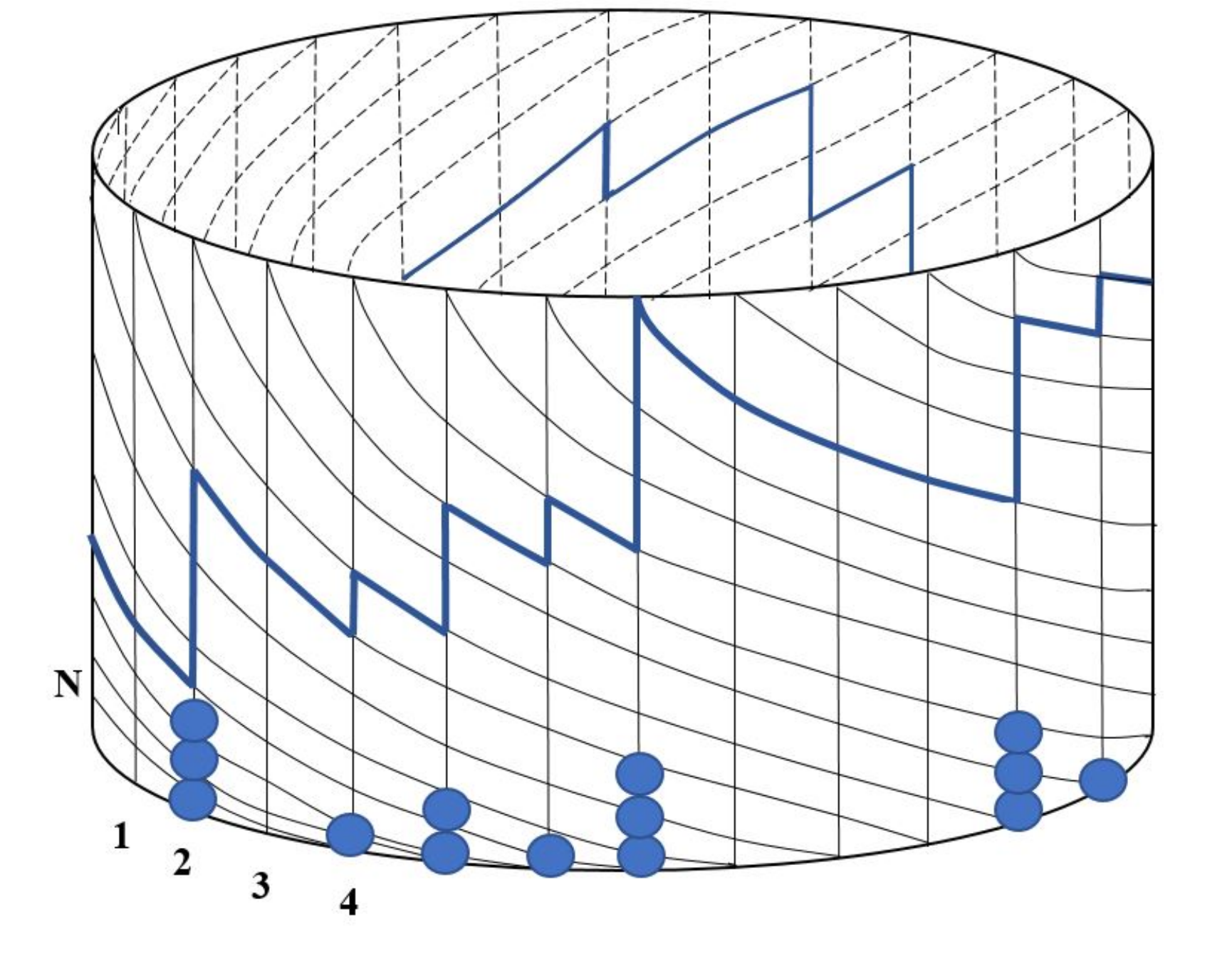} \caption{The mapping between configuration $\boldsymbol{n}=(0,3,0,1,2,1,3,0,0,0,3,1,0,\dots,0)$
and an interface.}
\label{cylinder}
\end{figure}

In this section exploiting the relation of q-boson ZRP with an interface
growth model we discuss the asymptotic limit of the announced exact
formulas in context of the KPZ-EW universality. The $q$-boson zero
range process on the periodic lattice can be mapped onto a growing
interface on a cylinder $\mathbb{R}\times[0,N]$. For $x\in[0,N]$
and time $t$ we define a piece-wise constant height function $h(x,t)$,
which experiences a jump 
\begin{equation}
h(x+0,t)-h(x-0,t)=n_{x}(t)\label{def_h}
\end{equation}
at each integer coordinate $x=1,\dots N$ and is constant otherwise.
Periodicity of the particle system implies helicoidal boundary conditions
for the height (see Fig.\ref{cylinder}) 
\begin{equation}
h(x+N,t)=\rho N+h(x,t),
\end{equation}
where the particle density $\rho=p/N$ plays the role of the mean
tilt of the interface. We are now interested in the late-time behaviour
of the large system, implying that that the limit $t\to\infty$ is
taken first. The statistics of the interface height in this limit
is dictated by the KPZ-EW universality. We first discuss the aspects
of the universal behaviour, which can be extracted from the KPZ equation.
Then we match this picture with the asymptotic results obtained for
the interface associated with the q-boson ZRP. 

\subsection{KPZ equation and scaling hypotheses }

Let us consider the KPZ equation for the interface height $h(x,t)$
\begin{equation}
\frac{\partial h}{\partial t}=\nu\frac{\partial^{2}h}{\partial x^{2}}+\frac{\lambda}{2}\left(\frac{\partial h}{\partial x}\right){}^{2}+\eta(x,t),\label{eq:KPZ}
\end{equation}
 $\eta(x,t)$ is a Gaussian white noise with zero mean and covariance
\begin{equation}
\mathbb{E}\eta(x,t),\eta(x',t')=D\delta(x-x')\delta(t-t'),\label{eq:white noise}
\end{equation}
where $\nu,\lambda$ and $D$ are three parameters of the model. The
particular case $\lambda=0$ is referred to as EW equation. We choose
helicoidal boundary conditions $h(x+N,t)=h(x,t)+\rho N$ with the
tilt $\rho$ for it to agree with the mapping \eqref{def_h}. We are
interested in the statistics of interface height function in the large
time limit. In this limit the statistics should not depend on initial
conditions. Consider the quantity characterizing the fluctuations
of the interface height on the cylinder, its dispersion, 
\begin{equation}
W^{2}(N,t)=\left\langle h^{2}(x,t)\right\rangle _{c}\label{W},
\end{equation}
with a particular case of the flat initial conditions, which ensure
that this quantity do not depend on the coordinate $x$ at any time.
The dimensional analysis shows that the general dependence of $W^{2}(N,t)$
on its arguments is given in terms of the function of $\mathcal{W}(x,y)$
of two dimensionless variables. 
\begin{equation}
W^{2}(N,t)=\frac{\nu^{2}}{\lambda^{2}}\mathcal{W}\left(\frac{\lambda^{2}DN}{\nu^{3}},\frac{\lambda^{4}D^{2}t}{\nu^{5}}\right).\label{eq: G}
\end{equation}
To make further predictions about its asymptotic limits one needs
to draw additional scaling arguments. We expect that at large time
under a specific mutual scaling this quantity interpolates between
the size independent small time regime, where $W^{2}(N,t)\asymp t^{2\alpha}$,
and the diffusive late time regime, $W^{2}(N,t)\simeq\Delta^{h}(N)t$
with the height diffusion coefficient $\Delta^{h}(N)$ depending on
$N$. The latter behaviour is due to the fact that at large time the
motion of a point of the interface is dominated by the motion of the
interface center of mass
\[
\bar{h}_{N}(t)=\frac{1}{N}\int_{0}^{N}h(x,t)dx,
\]
in view of which we could equally discuss the quantity 
\begin{equation}
W_{c}^{2}(N,t)=\left\langle \bar{h}^{2}(t)\right\rangle _{c}\label{W_c}
\end{equation} characterizing the fluctuations of the center mass of the interface. Though  the  quantities  $W(N,t)$ and   $W_c(N,t)$ defined by (\ref{W}) and (\ref{W_c}) respectively  are manifestly different, they asymptotically coincide in the  limit under consideration. In view of this, we do not distinguish between them in what follows. 
The small and large time regimes are expected to meet at the characteristic
relaxation time $\tau(N)\asymp N^{z}$. In particular, when the temporal
and spacial scales are related by this scaling we require the functional
form (\ref{eq: G}) to be consistent with the Family-Viseck-like ansatz
\citep{FV}
\begin{equation}
W^{2}(N,t)=t^{2\alpha}\Phi(t/\tau(N)).
\end{equation}
with the scaling function $\Phi(x),$ that behaves as $\Phi(x)\to\mathrm{const}$
and $\Phi(x)\asymp x^{1-2\alpha}$ as $x\to0$ and $x\to\infty$ respectively.
To connect this ansatz to (\ref{eq: G}) we introduce the function
$\mathcal{H}(x,y)$
\begin{equation}
\mathcal{W}(x,y)=y^{2\alpha}\mathcal{H}(x,y).\label{eq:H}
\end{equation}
When $y\asymp x^{z}\to\infty$, $\mathcal{H}(x,y)$ is supposed to
take the scaling form
\begin{equation}
\lim_{L\to\infty}\mathcal{H}(Lx,L^{z}y)=\Phi(y/x^{z}).\label{eq:H-FV}
\end{equation}
Substituting (\ref{eq:H},\ref{eq:H-FV}) into (\ref{eq: G}) we obtain
for the dependence of $W^{2}(N,t)$ on the parameters 
\begin{equation}
W^{2}(N,t)\simeq\kappa_{KPZ}(D/2\nu)^{\frac{3}{2}}|\lambda|N^{-\frac{1}{2}}t,\quad t\gg N^{3/2},\ N\rightarrow\infty,\label{eq:W KPZ}
\end{equation}
and 
\begin{equation}
W^{2}(N,t)\simeq\kappa_{EW}\frac{Dt}{N},\quad t\gg N^{2},\ N\rightarrow\infty\label{eq: W EW}
\end{equation}
for KPZ and EW classes respectively, where $\kappa_{KPZ}$ and $\kappa_{EW}$
are the universal dimensionless constants specific for given universality
class. These constants can be obtained from exact solutions. The latter
\begin{equation}
\kappa_{EW}=1\label{eq:kappa_EW}
\end{equation}
 follow directly from the EW equation, which suggests that $\bar{h}(t)$
moves as a Brownian motion with the dispersion $D$ at any time $t$.
The former 
\begin{equation}
\kappa_{KPZ}=\frac{\sqrt{\pi}}{4}
\end{equation}
was first obtained from the exact solution of the TASEP \citep{DerLeb}
and conjectured to be universal for the whole KPZ universality class. 

Note that there are two different functions $\Phi(x)$ for EW and
KPZ classes. The crossover between their late time asymptotics is
given in terms of yet another scaling function $\mathcal{F}(x,y),$
\[
\mathcal{W}(x,y)=y\mathcal{F}(x,y/x^{2})/x.
\]
The crossover describes the late time stage of the evolution in the
diffusive scale $t/N^{2}\to\infty$ under the scaling $\lambda\asymp1/\sqrt{N},\quad N\to\infty.$
Then, the height dispersion 
\begin{equation}
\lim_{t\rightarrow\infty}\frac{W^{2}(N,t)}{t}=\frac{D}{N}\mathcal{F}(g,\infty),\quad g=\frac{\lambda^{2}DN}{\nu^{3}}.\label{sc_fun}
\end{equation}
is given in terms of $\mathcal{F}(g,\infty),$ which is conjecturally
the universal crossover function. The candidate for this function
\begin{equation}
\mathcal{F}(g,\infty)=\frac{\sqrt{g}}{2\sqrt{2}}\int_{0}^{+\infty}\ \frac{y^{2}e^{-y^{2}}}{\tanh((\sqrt{g}/\sqrt{32})y)}dy.\label{eq:F(g,infty)}
\end{equation}
was first was obtained in \citep{DerMal} as a scaling limit of the
exact diffusion constant at the weak asymmetry. 

Below we are going to test these conjectures against the asymptotic
results obtained from the q-boson ZRP. To this end, we, however, still
need a way of identification of the parameters of the model with those
of the KPZ equation.

\subsection{Dimensionful invariants and asymptotic results.\label{subsec:Dimensionful-invariants-and}}

The way of identification of the model dependent constants in the
KPZ universality class was proposed in \citep{KMH,AF}. It is based
on the observation that the parameters $A=D/2\nu$ and $\lambda$
are stable with respect to scale transformation 
\begin{equation}
x\to bx,t\to b^{z}t,h\to b^{\zeta}h,\label{eq:scaling transfom}
\end{equation}
with $z=3/2$ and $\zeta=1/2$, which together with the corresponding
transformation of $D,\lambda$ and $\nu$ leaves the KPZ equation
invariant. It was then conjectured that the dimensionful model-dependent
constants within the universal functions must appear as a combination
of these two parameters. It is indeed the case in (\ref{eq:W KPZ}),
which suggests
\begin{equation}
W^{2}(N,t)\simeq\kappa_{KPZ}A{}^{\frac{3}{2}}|\lambda|N^{-\frac{1}{2}}t,\label{eq:W(A,lambda)}
\end{equation}
and is conjecturally true for all the systems of KPZ class.

The quantities $A$ and $\lambda$ can be expressed in terms of the
characteristics of the stationary state of the process in the infinite
system. Specifically the non-linearity coefficient $\lambda$ is the
second derivative 
\begin{equation}
\lambda=\frac{\partial^{2}v_{\infty}^{h}}{\partial\rho^{2}},\label{def_lambda}
\end{equation}
of the infinite system limit $v_{\infty}^{h}=\lim_{N\to\infty}v_{N}^{h}$
of the mean interface velocity 
\[
v_{N}^{h}=\lim_{t\to\infty}\frac{h(x,t)}{t}
\]
 with respect to the tilt, while the constant $A$ can be defined
as the amplitude of the two-point interface height covariance 
\begin{equation}
\lim_{N\to\infty}\lim_{t\rightarrow\infty}\langle(h(x,t)-h(y,t))^{2}\rangle_{c}=A|x-y|.\label{def_A}
\end{equation}
Yet another quantity, which can be expressed in terms of the dimensionful
invariants in a universal way is the leading finite size correction
to the interface velocity 
\begin{equation}
\text{\ensuremath{\lim_{N\to\infty}N^{-1}(v_{N}^{h}-v_{\infty}^{h})=-\frac{A\lambda}{2}.}}\label{eq:fss}
\end{equation}

To apply these formulas to the interface associated with the q-boson
ZRP we remind that from the ZRP-interface mapping introduced in the
beginning of the section the number of particles $y_{i}(t)$ passed
through the bond $(i,i+1)$ by time $t$ is the increase of the height
$(h(x,t)-h(x,0))$ for $x\in(i,i+1)$. Hence the height dispersion
is 
\[
W(N,t)=\left\langle y_{i}^{2}\right\rangle _{c}\eqsim\frac{\Delta t}{N^{2}},\quad t\to\infty.
\]
The interface velocity is nothing but the mean particle current
\begin{equation}
v_{N}^{h}=j_{N}\label{eq: v_N^h}
\end{equation}
 and the two-point height covariance is the variance of the number
of particles between the reference points
\[
\lim_{N\rightarrow\infty}\lim_{t\rightarrow\infty}\langle(h(x,t)-h(y,t))^{2}\rangle_{c}=\left\langle \left(\sum_{x<i<y}n_{i}\right)^{2}\right\rangle _{c}.
\]

The two latter quantities can alternatively be obtained as averages
over the stationary state. To this end we evaluate the contour integrals
from the subsection \ref{subsec:Stationary-state-and} asymptotically
in the thermodynamic limit 
\begin{equation}
p\rightarrow\infty,\quad N\rightarrow\infty,\quad\frac{p}{N}=\rho.\label{thermodyn_limit}
\end{equation}
 using the saddle point approximation. The details of the method will
be given in section \ref{sec:Asymptotic-analysis}. Here we place
the results of the calculation. They are given in terms of the derivatives
\begin{eqnarray*}
h_{k} & = & (z\partial_{z})^{k}h(z)|_{z=z^{*}}
\end{eqnarray*}
 of the function 
\begin{eqnarray*}
h(z) & = & \ln F(z)-\rho\ln(z)
\end{eqnarray*}
 at the critical point $z^{*}$ of the integrand, given by the smallest
positive solution of the equation 
\begin{equation}
z^{*}(\ln F(z^{*}))'=\rho\label{def_z*}
\end{equation}
with the function $F(z)$ defined in (\ref{qexp}). In the statistical
physics language the latter equation is the standard relation between
the density $\rho$ and fugacity $z^{*}$, while the function $\ln F(z)$
plays the role of the grand-canonical free energy. 

The evaluation of the integral (\ref{eq: def_Z_sum}) asymptotically
yields

\begin{equation}
Z(N,p)=\frac{e^{Nh_{0}}}{\sqrt{2\pi Nh_{2}}}\left[1+\frac{1}{2N}\left(\frac{h_{4}}{4h_{2}^{2}}-\frac{5h_{3}^{2}}{12|h_{2}|^{3}}\right)+O(N^{-2})\right].\label{eq:Z asymp}
\end{equation}
For the current in the infinite system we obtain 
\begin{equation}
j_{\infty}(\rho)=z^{*}\label{eq:j_infty}
\end{equation}
together with the finite size correction 
\begin{equation}
\lim_{N\to\infty}N(j_{N}-j_{\infty})=\frac{1}{2}\left(\frac{h_{3}}{h_{2}^{2}}-\frac{1}{|h_{2}|}\right)\label{eq: current fss}
\end{equation}
For the non-linearity coefficient we have
\begin{equation}
\lambda=\frac{\partial^{2}j_{\infty}}{\partial\rho^{2}}=\frac{z^{*}}{h_{2}}\left(\frac{1}{|h_{2}|}-\frac{h_{3}}{h_{2}^{2}}\right).\label{lambda_q_boson}
\end{equation}
where we used that $dz^{*}/d\rho=z^{*}/h_{2},$ which follows from
\eqref{def_z*}. 

The variance of the particle number between two given points is the
sum of variances of single site occupancies in this interval due to
decorrelation of the occupancies in the thermodynamic limit 
\begin{eqnarray}
\left\langle \left(\sum_{x<i<y}n_{i}\right)^{2}\right\rangle _{c} & = & \sum_{x<i<y}\langle n_{i}^{2}\rangle_{c}\eqsim h_{2}\left|x-y\right|,\quad\left|x-y\right|\to\infty,
\end{eqnarray}
which yields 
\begin{equation}
A=h_{2}.\label{eq: A=00003Dh_2}
\end{equation}
We thus have expressed the dimensionful invariants $A$ and $\lambda$
in terms of characteristics of the stationary state of an infinite
system. It also can be checked by a direct computation that these
definitions are consistent with the formula (\ref{eq:fss}) applied
to the finite size corrections (\ref{eq: current fss}) of the particle
current $j_{N}$. 

The results of substitution of the formulas \eqref{lambda_q_boson},\eqref{eq: A=00003Dh_2}
into \eqref{eq:W(A,lambda)} are to be compared with the following
result of the asymptotic analysis of exact formula \eqref{Answer}
carried out in section \ref{sec:Asymptotic-analysis}.

\textbf{Result 2.}
\begin{equation}
\lim_{N\to\infty}\frac{\Delta}{N^{\frac{3}{2}}}=\frac{\sqrt{\pi}}{8\sqrt{h_{2}}}\left(\frac{\phi_{1}h_{3}}{|h_{2}|}-\phi_{2}\right),\label{lambda_2_KPZ_regime-1}
\end{equation}
where $\phi_{k}=(z\partial_{z})^{k}\phi(z)|_{z=z^{*}}$ with $\phi(z)$
defined in \ref{def_phi}. Surprisingly, up to the error of order
of $O(1/N)$ this formula can also be recast in terms of the stationary
state observables of the original system and of the similar system
with twice larger size and number of particles,
\begin{equation}
\frac{\Delta}{N^{2}}=\frac{Z(2N,2p)}{Z(N,p)^{2}}\left(j_{N}(\rho)-j_{2N}(\rho)\right)+O\left(1/N\right).\label{eq:Delta via fss}
\end{equation}
 Specifically, its leading asymptotics is given in terms of the universal
finite size corrections to the current. By virtue of (\ref{eq:Z asymp},
\ref{eq: current fss}) and (\ref{eq:fss},\ref{eq: v_N^h}) this
is the same as (\ref{eq:W(A,lambda)}). It is an interesting question
whether this formula is meaningful in a more general context. 

The formulas (\ref{lambda_2_KPZ_regime-1},\ref{eq:Delta via fss})
also agree with the universal scaling expression of the large deviation
function obtained in \citep{P2003}, where the dimensionful constants
were expressed in terms of the function $g_{\nu}(z)=\sum_{i=1}^{\infty}z^{i}/(1-\nu^{i})$
related to the quantities under consideration by $g_{q^{\pm1}}(\pm(1-q^{\pm1})z)$
$=\pm z(\ln F(z))'$ for $q\lessgtr1$ respectively. 

The second part of the asymptotic analysis is devoted to the crossover
regime. It corresponds to the scaling, in which the dimensionless
variable $g=\lambda^{2}DN\nu^{-3}$ from (\ref{sc_fun}) stays finite
as $N\to\infty$. This can be realized by taking 
\[
q=e^{-\frac{\alpha}{\sqrt{N}}}.
\]
In this case 
\[
j_{N}=\rho-\frac{\alpha\rho^{2}}{2\sqrt{N}}+O\left(\frac{1}{N}\right)
\]
and hence
\[
\lambda\simeq-\frac{\alpha}{\sqrt{N}}.
\]
Thus $\alpha$ varying from zero to infinity brings the system from
EW to KPZ universality class. Correspondingly the parameters, $D$
and $\nu$ should be given the limiting values they take in the EW
limit. Note that like the parameters $A$ and $\lambda$ are the dimensionful
invariants of the KPZ class, the parameters $D$ and $\nu$ are invariant
with respect to the scaling transformation (\ref{eq:scaling transfom})
with exponents $\zeta=1/2$ and $z=2$ that leave the EW equation
invariant. Thus, conjecturally these quantities can be ascribed to
any model in the EW universality class. In our case, when $q=1$ we
use  (\ref{eq:q=00003D1 curren and diff coef}) and (\ref{eq: W EW},\ref{eq:kappa_EW})
to show that 
\[
D=\rho.
\]
 To find $\nu$ we note that $A=D/(2\nu)=h_{2}=\rho$ in the same
limit, form where we conclude that 
\[
\nu=\frac{1}{2}
\]
and $g=\text{\ensuremath{8\rho\alpha^{2}}}$. 

With such a defined $g$ the result of the asymptotic analysis of
the exact formula \eqref{Answer} of the diffusion coefficient in
the crossover regime is as follows. 

\textbf{Result 3.}

\begin{equation}
\lim_{N\to\infty}\frac{\Delta}{N}=\rho\mathcal{F}(g,\infty),\label{Sc_limit<1}
\end{equation}
which is exactly matches with the conjectured expression (\ref{sc_fun})
with the universal scaling function $\mathcal{F}(g,\infty)$ from
(\ref{eq:F(g,infty)}).

\section{Exact formulas for scaled current cumulants}

\subsection{Bethe ansatz and T-Q equation}

The purpose of this section is to derive the integral representations
for the first two derivatives of the largest eigenvalue of the operator
$\mathcal{L_{\gamma}}$. The operator $\mathcal{L_{\gamma}}$ for
$q$-boson zero range process can be diagonalized using the Bethe
Ansatz \citep{SW,P2003}. The eigenvectors and eigenvalues are parameterized
by $p$ complex numbers $x_{1},\dots,x_{p}$ given by the roots of
Bethe ansatz equations (BAE) 
\begin{equation}
e^{\gamma N}(1-x_{i})^{-N}=(-1)^{p-1}\prod_{j=1}^{p}\frac{x_{i}-qx_{j}}{x_{j}-qx_{i}}.\label{BAE}
\end{equation}
A particular solution corresponds to a particular eigenvector, and
the corresponding eigenvalue has the form 
\begin{equation}
\lambda(\gamma)=-\sum_{i=1}^{p}x_{i}.
\end{equation}
We would like to find a particular solution of BAE corresponding to
the Perron-Frobenius eigenvalue satisfying $\lambda(0)=0$. The corresponding
eigenvector is known to be translationally invariant, that is to say
that for the solution of BAE we have 
\begin{equation}
\prod_{i=1}^{p}(1-x_{i})=e^{\gamma p}.\label{trans_inv}
\end{equation}
Let us introduce a polynomial $Q(x)$ of degree $p$, with roots from
the solution of BAE 
\[
Q(x)=\prod_{i=1}^{p}(x-x_{i}).
\]
Rewriting the system (\ref{BAE}) in terms of $Q(x)$ we notice that
the Bethe roots are also zeroes of the polynomial 
\[
e^{\gamma N}Q(qx)+q^{p}(1-x)^{N}Q(x/q).
\]
This fact suggests that the latter polynomial is divisible by $Q(x)$.
Hence we have the polynomial TQ-relation 
\begin{equation}
T(x)Q(x)=e^{\gamma N}Q(qx)+q^{p}(1-x)^{N}Q(x/q).\label{TQ}
\end{equation}
between the polynomial $Q(x)$ of degree $p$ and yet another unknown
polynomial $T(x)$ of degree $N$. Together with the condition 
\begin{equation}
Q(1)=e^{\gamma p}\label{Q(1)}
\end{equation}
following from \eqref{trans_inv} it determines the functional equation
for $Q(x)$ corresponding to the largest eigenvalue.

Once we know $Q(x)$, the eigenvalue is 
\begin{equation}
\lambda(\gamma)=\left.\frac{1}{(p-1)!}\frac{d^{p-1}Q(x)}{dx^{p-1}}\right|_{x=0}.\label{ddQ}
\end{equation}
We are going to solve relation (\ref{TQ}) perturbatively in powers
of $\gamma$ in the vicinity of $\gamma=0$. To this end, we assume
the following expansions 
\begin{eqnarray*}
Q(x) & = & Q_{0}(x)+\gamma Q_{1}(x)+\gamma^{2}Q_{2}(x)+\dots\text{,}\\
T(x) & = & T_{0}(x)+\gamma T_{1}(x)+\gamma^{2}T_{2}(x)+\dots,\\
\lambda(\gamma) & = & \gamma\lambda_{1}+\gamma^{2}\lambda_{2}+\dots.
\end{eqnarray*}

The T-Q relation is equivalent to the system of equations for the
polynomials $T_{k}(x)$ and $Q_{k}(x)$ which can be solved order
by order. The leading orders of $Q(x)$ and $T(x)$ follow directly
from (\ref{TQ}).
\begin{equation}
Q_{0}(x)=x^{p},\ \ \ T_{0}(x)=q^{p}+(1-x)^{N}.\label{0 order}
\end{equation}
The first two coefficients of the eigenvalue are of our interest,
being related to the scaled cumulants of integrated current.
\[
\text{\ensuremath{\lambda_{1}=J,\,\,}\ensuremath{\lambda_{2}=\frac{\Delta}{2}}}
\]
To find them we solve T-Q relation in the first and second orders. 

\subsection{First order calculation}

The equation \eqref{TQ} in the first order in $\gamma$ looks as
follows 
\begin{equation}
T_{0}(x)Q_{1}(x)+T_{1}(x)Q_{0}(x)=Q_{1}(qx)+NQ_{0}(qx)+q^{p}(1-x)^{N}Q_{1}(x/q).\label{TQ 1 order}
\end{equation}
As $\deg Q_{1}(x)\leq p-1$, it is enough to solve this equation $\mathrm{mod}\,\,x^{p}$.
Due to \eqref{0 order} it is then reduced to 
\begin{equation}
(q^{p}+(1-x)^{N})Q_{1}(x)=Q_{1}(qx)+q^{p}(1-x)^{N}Q_{1}(x/q)\,\,\mathrm{mod}\,\,x^{p}.
\end{equation}
Introducing 
\begin{equation}
B_{1}(x)=q^{p}Q_{1}(x/q)-Q_{1}(x)\label{def_of_B_1}
\end{equation}
we obtain 
\begin{equation}
B_{1}(qx)=(1-x)^{N}B_{1}(x)\,\,\mathrm{mod}\,\,x^{p}.\label{B_1 eq}
\end{equation}

The cases $|q|<1$ and $|q|>1$ should be treated separately. In both
cases we first construct the solution $\tilde{B}_{1}(x)$ of the difference
equation regular at $x=0$, which is globally a meromorphic function
in the first case and entire function in the second. For that purpose
we iterate equation (\ref{B_1 eq}) obtaining for both $|q|<1$ and
$|q|>1$ the solution 
\begin{eqnarray*}
\tilde{B}_{1}(x) & = & \tilde{B}_{1}(0)F\left(x/\left(1-q\right)\right){}^{N},
\end{eqnarray*}
where $F(z)$ was defined in (\ref{qexp}). Having written the formula
for $\tilde{B}_{1}(x)$ in the same way for all values of $q\neq1$
we can proceed further without referring to the value of $q$.

The first $p$ terms of the expansion $\tilde{B}_{1}(x)$ at the origin
give the desired polynomial $B_{1}(x)=\sum_{i=0}^{p-1}b_{i}x^{i}$.
Its coefficients define the polynomial $Q_{1}(x)=\sum_{i=0}^{p-1}q_{i}x^{i}$
due to the relation 
\begin{equation}
b_{i}=(q^{p-i}-1)q_{i}.\label{bviaq}
\end{equation}
which follows from (\ref{def_of_B_1}). Then one can write down the
relation between the polynomials $Q_{1}(x)$ and $B_{1}(x)$ in an
integral form 
\begin{eqnarray}
Q_{1}(x) & = & \sum_{i=0}^{p-1}\frac{b_{i}x^{i}}{q^{p-i}-1}=\sum_{i=1}^{p}\frac{b_{p-i}x^{p-i}}{q^{i}-1}\label{eq:Q_1}\\
 &  & =-x^{p}\oint\frac{B_{1}(z)}{z^{p+1}}\sum_{i=1}^{\infty}\frac{(z/x)^{i}}{1-q^{i}}\frac{dz}{2\pi\mathrm{i}}\nonumber 
\end{eqnarray}
Here the integration contour is a simple anticlockwise loop around
the origin $z=0$, which must be the only singularity inside the contour.

Then, it will be convenient to express everything in terms of $F(x)$.
It turns out that the results expressed in terms of this function
look the same for all values of $q\neq1$. This is in particular due
to the fact that for the logarithmic derivative of $F(z)$ we have
\begin{equation}
(\ln F(x))'=\sum_{k=0}^{\infty}\frac{x^{k}(1-q)^{k+1}}{1-q^{k+1}}\label{ln'F}
\end{equation}
irrespective of whether $|q|<1$ or $|q|>1$. This expression, however,
is valid in the domain of the series convergence, $|(1-q)x|<1$ in
the former case and $|(1-q)x/q|<1$ in the latter.

Substituting $\tilde{B}_{1}(x)$ to \eqref{eq:Q_1} instead of $B_{1}(x)$,
which does not affect the result of integration as only the first
$p$ terms of the expansion of these functions near the origin contribute
to the integral, we obtain 
\begin{eqnarray}
Q_{1}(x)=-\frac{\tilde{B}_{1}(0)x^{p-1}}{(1-q)^{p}}\oint\frac{F(z)^{N}}{z^{p}}(\ln F\left(\frac{z}{x}\right))'\frac{dz}{2\pi\mathrm{i}},\label{Q_1(x)}
\end{eqnarray}
where the integration variable was scaled by $(1-q)^{-1}$. The constant
$B_{1}(0)$ is defined from \eqref{Q(1)} where we substitute the
integral form of $Q_{1}(x)$.

For further purposes we list the resulting representations for $Q_{1}(x)$,$B_{1}(x)$
\begin{equation}
\tilde{B}_{1}(x)=-\frac{N(1-q)^{p}F(x/(1-q))^{N}}{Z(N,p)},\label{int_form_of_B1}
\end{equation}
\begin{equation}
Q_{1}(x)=\frac{Nx^{p-1}}{Z(N,p)}\oint\frac{F(z)^{N}}{z^{p}}(\ln F\left(\frac{z}{x}\right))'\frac{dz}{2\pi\mathrm{i}}.\label{Q1(x) int form}
\end{equation}
The the integration contours in the latter formula satisfy $|(1-q)z|<|x|$
and $|(1-q)z|<|qx|$ for $|q|<1$ and $|q|>1$, respectively.

We are in a position to write the expression for $\lambda_{1}$. 
\begin{eqnarray}
\lambda_{1} & = & \left.\frac{1}{(p-1)!}\frac{d^{p-1}Q_{1}(x)}{dx^{p-1}}\right|_{x=0}=\frac{N}{Z(N,p)}\oint\frac{F(z)^{N}}{z^{p}}\frac{dz}{2\pi\mathrm{i}}=N\frac{Z(N,p-1)}{Z(N,p)},
\end{eqnarray}
which coincides with stationary state result.

\subsection{Second order calculation}

The second order equation is as follows 
\begin{eqnarray}
 &  & \frac{N^{2}}{2}Q_{0}(qx)+NQ_{1}(qx)+Q_{2}(qx)+q^{p}(1-x)^{N}Q_{2}(x/q)\nonumber \\
 &  & =Q_{1}(x)T_{1}(x)+Q_{2}(x)T_{0}(x)+Q_{0}(x)T_{2}(x)
\end{eqnarray}
with initial condition 
\begin{eqnarray}
\frac{p^{2}}{2}=Q_{2}(1)\label{Q_2(1)}
\end{eqnarray}
which comes from the second order of (\ref{Q(1)}). Repeating the
same reasoning as in case of the first correction we write
\begin{eqnarray}
NQ_{1}(qx)+Q_{2}(qx)+q^{p}(1-x)^{N}Q_{2}(x/q)=\\
=Q_{1}(x)T_{1}(x)+Q_{2}(x)(q^{p}+(1-x)^{N})\quad\mathrm{mod}\,\,x^{p}.\nonumber 
\end{eqnarray}
Introducing 
\begin{equation}
B_{2}(x)=q^{p}Q_{2}(x/q)-Q_{2}(x)
\end{equation}
we obtain 
\begin{eqnarray}
-B_{2}(qx)+(1-x)^{N}B_{2}(x)=Q_{1}(x)T_{1}(x)-NQ_{1}(qx)\quad\mathrm{mod}\,\,x^{p}.\label{B_2 eq}
\end{eqnarray}
Let us denote $f(x)$ the right hand side of this equation with $T_{1}(x)$
found from \eqref{TQ 1 order}, 
\begin{eqnarray}
T_{1}(x)=Nq^{p}+x^{-p}[(1-x)^{N}B_{1}(x)-B_{1}(qx)].
\end{eqnarray}
To solve the equation \eqref{B_2 eq} we rewrite it in the standard
form with $q$-difference operator defined by 
\begin{equation}
D_{q}a(x)=\frac{a(qx)-a(x)}{(q-1)x},
\end{equation}
which yields
\begin{equation}
D_{q}B_{2}(x)=\frac{-f(x)}{x(q-1)}+\frac{(1-x)^{N}-1}{x(q-1)}B_{2}(x)\mathrm{\,\,mod}\,\,x^{p}\label{eq for B_2}
\end{equation}
This is the first order linear inhomogeneous q-difference equation
with non-constant coefficients. Depending on whether $|q|<1$ or $|q|>1$
corresponding homogeneous equation is solved by iterations with decreasing
of increasing powers of $q$, so that the final expressions slightly
differ. Below we show the calculation for $|q|<1$ in detail. Calculations
for the case $|q|>1$ are analogous and will be omitted. 

The solution of (\ref{eq for B_2}) can be obtained by the variation
of constant method

\begin{eqnarray}
\tilde{B_{2}}(x)=S\left(F\left(x/(1-q)\right)\right)^{N}+\left(F\left(x/(1-q)\right)\right)^{N}\sum_{i=0}^{\infty}\frac{f(q^{i}x)}{F^{N}\left(q^{i+1}x/(1-q)\right)},
\end{eqnarray}
where $S$ is the constant of integration to be defined from \eqref{Q_2(1)}.
The coefficients of polynomials $Q_{2}$ and $B_{2}$ are related
by (\ref{bviaq}). Then, one has the integral representation for $Q_{2}(x)$
similar to (\ref{Q1(x) int form}) 
\begin{equation}
Q_{2}(x)=-\frac{x^{p-1}}{(1-q)^{p}}\oint\frac{F^{N}(z)}{z^{p}}\left(S+\sum_{i=0}^{\infty}\frac{f(q^{i}(1-q)z)}{F(q^{i+1}z)^{N}}\right)(\ln F\left(\frac{z}{x}\right))'\frac{dz}{2\pi\mathrm{i}}
\end{equation}
The constant $S$ is found from the initial condition 
\begin{equation}
S=-\frac{pN(1-q)^{p}}{2Z(N,p)}-\frac{N}{pZ(N,p)}\oint\frac{F^{N}(z)}{z^{p}}\sum_{i=0}^{\infty}\frac{f(q^{i}(1-q)z)}{F(q^{i+1}z)^{N}}(\ln F(z))'\frac{dz}{2\pi\mathrm{i}}
\end{equation}
Finally we have 
\begin{eqnarray*}
Q_{2}(x) & = & \frac{pNZ(N,p-1)}{2Z(N,p)}+\frac{x^{p-1}}{(1-q)^{p}}\oint\frac{F^{N}(z)}{z^{p}}\biggl[\sum_{i=0}^{\infty}\frac{f(q^{i}(1-q)z)}{F(q^{i+1}z)^{N}}\\
 & \times & \left(\frac{N}{pZ(N,p)}(\ln F(z))'\oint\frac{F^{N}(\tilde{z})}{\tilde{z}^{p}}(\ln F\left(\frac{\tilde{z}}{x}\right))'\frac{d\tilde{z}}{2\pi i}-(\ln F\left(\frac{z}{x}\right))'\right)\biggr]\frac{dz}{2\pi\mathrm{i}}.
\end{eqnarray*}
We are interested in the highest coefficient of $Q_{2}(x)$. To this
end, we divide the last expression by $x^{p-1}$ and send $x$ to
infinity. Only the free term of $(\ln F(z/x))'$ equal to unity remains.
\begin{eqnarray}
\lambda_{2} & = & \frac{pNZ(N,p-1)}{2Z(N,p)}+\frac{1}{(1-q)^{p}}\sum_{i=0}^{\infty}\oint\frac{F^{N}(z)}{z^{p}}\phi(z)\frac{f(q^{i}(1-q)z)}{F(q^{i+1}z)^{N}}\frac{dz}{2\pi\mathrm{i}}\label{Prelim_lambda2}
\end{eqnarray}
with $\phi(x)$ defined in (\ref{def_phi}).

The next step is to simplify this result by substituting $f(x)$ and
the integral forms of polynomials $B_{1}(x)$ and $Q_{1}(x)$ that
is done step by step in \ref{sec:Derivation-of-equation}. As a result
we arrive at the following sum of triple integrals
\begin{eqnarray}
\lambda_{2} & = & \frac{pJ}{2}+\frac{N^{2}}{Z(N,p)^{2}}\nonumber \\
 & \times & \oint\frac{dz\phi(z)}{2\pi iz}\biggl[\oint\limits _{|y|<|z|}\frac{dy}{2\pi i}\frac{F^{N}(y)}{y^{p}}\oint\limits _{|z|<|t|}\frac{dt}{2\pi i}\frac{F^{N}(t)}{t^{p}}\frac{(\ln F\left(\frac{y}{(1-q)z}\right))'}{(t-z)(1-q)}\nonumber \\
 & + & z\sum_{i=0}^{\infty}\oint\limits _{|y|<|q^{i+1}z|}\frac{dy}{2\pi i}\frac{F^{N}(y)q^{(i+1)(p-1)}}{y^{p}(z-q^{-i-1}y)}\oint\limits _{|q^{i+1}z|<|t|}\frac{dt}{2\pi i}\frac{F^{N}(t)}{t^{p}}\frac{F_{0,i+1}^{N}(z)}{t-q^{i+1}z}\biggr],\label{lambda2 before integration on z}
\end{eqnarray}
where $J$ is the mean integrated current given by (\ref{eq:J}) and
\begin{equation}
F_{0,i}(z):=\frac{F(z)}{F(q^{i}z)}.
\end{equation}
To obtain the final formula we notice that the integral in $z$ can
be evaluated by the residue calculus. All the poles inside the contour
contributing to the first term in square brackets of \eqref{lambda2 before integration on z}
are those coming from $(\ln F\left(y/((1-q)z)\right))'$, i.e. $z=q^{i}y,\quad i=0,1,\dots$
A summand of the second term with number $i$ receives the contribution
from the only pole $z=q^{-i-1}y$. After variable change $\tilde{y}=q^{-i-1}y$
we obtain 
\begin{eqnarray*}
\lambda_{2} & = & \frac{pJ}{2}+\frac{N^{2}}{Z(N,p)^{2}}\sum_{i=0}^{\infty}\oint\limits _{|y|<|t|}\frac{dy}{2\pi i}\frac{F^{N}(y)}{y^{p}}\oint\frac{dt}{2\pi i}\frac{F^{N}(t)}{t^{p}}\frac{q^{i}\phi(yq^{i})}{t-yq^{i}}.\\
 & + & \frac{N^{2}}{Z(N,p)^{2}}\sum_{i=0}^{\infty}\oint\limits _{|y|<|t|}\frac{dt}{2\pi i}\frac{F^{N}(t)}{t^{p}}\oint\frac{dy}{2\pi i}\frac{F^{N}(y)}{y^{p}}\frac{\phi(y)}{t-yq^{i+1}}
\end{eqnarray*}
 which is a half of the expression for $\Delta$ announced in \eqref{Answer}
in the case $|q|<1$. The case $q>1$ derived in a similar way is
obtained from this formula by replacing $q\to1/q$.

\section{Asymptotic analysis\label{sec:Asymptotic-analysis}}

Let us move to the asymptotic analysis of the exact formulas in the
thermodynamic limit
\[
N\to\infty,p\to\infty,p/N\to\rho.
\]
The main ingredient of the analysis is the evaluation of integrals
of the form 
\begin{equation}
\oint\frac{dt}{2\pi i}\frac{F^{N}(t)}{t^{p+1}}g(z)=\oint\frac{dz}{2\pi iz}e^{Nh(z)}g(z),\label{def_I}
\end{equation}
and its two-dimensional analogues in the saddle point approximation.
Here $h(z)=\ln F(z)-\rho\ln(z)$ was defined in (\ref{def_h}) and
$g(z)$ is an arbitrary function analytic on the integration contour.
The main contribution to this integral comes from a critical point
$z^{*}$ of $h(z)$ given by the solution of 
\[
h'(z^{*})=0.
\]
\begin{figure}
\noindent \centering{}\includegraphics[width=0.7\linewidth]{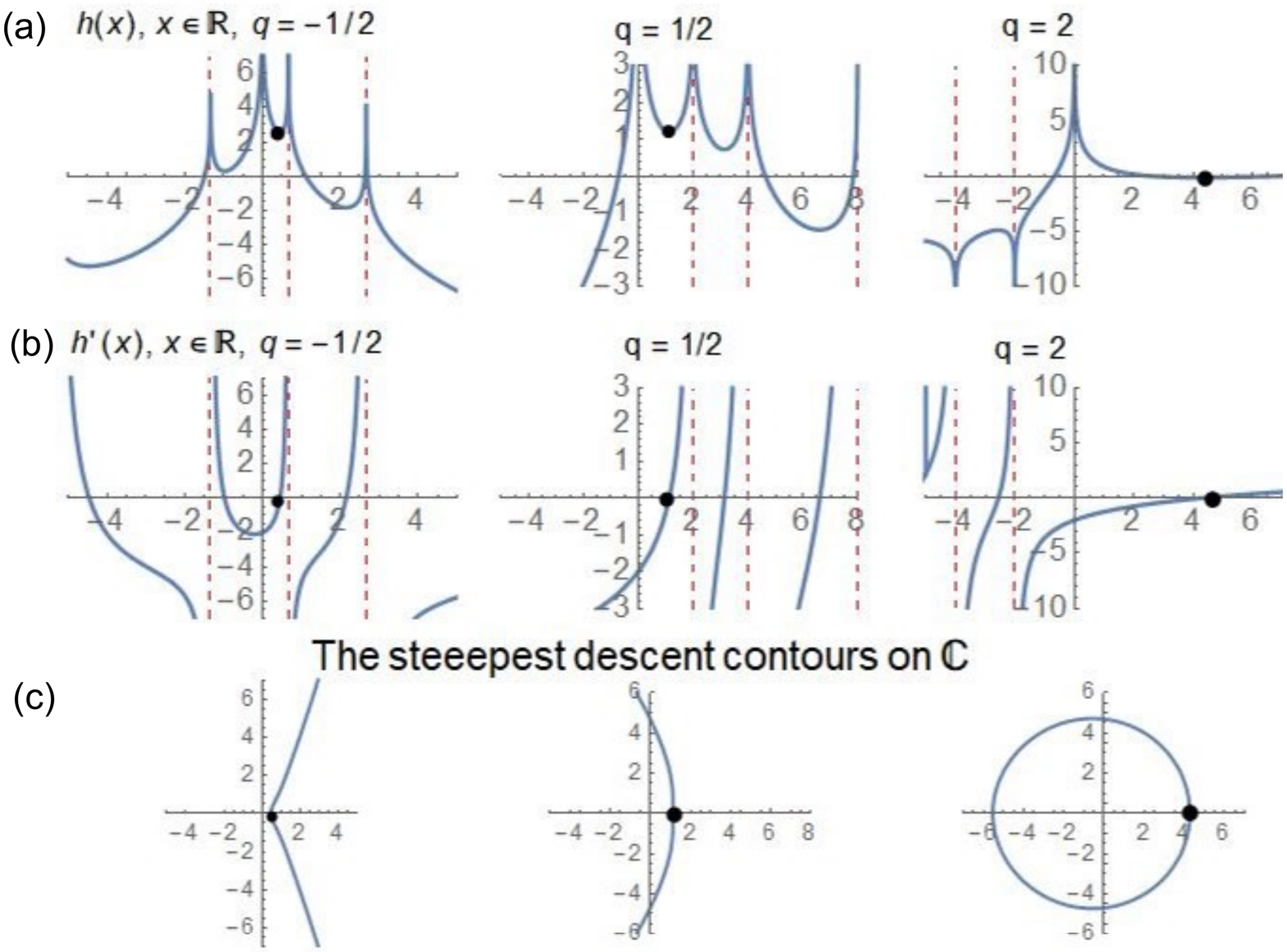}
\caption{The graphs of $\Re h(z)$ (a) and of $h'(z)$ (b) for different values
of $q$. The density is $\rho=2$ in all cases. The bottom row (c)
shows corresponding contours of steepest descent in  the complex plane.}
\label{fig: contours}
\end{figure}

There are many solutions of this equation, all being on the real axis
(see fig.\ref{fig: contours}). It can be seen from the figure that
the maximal contribution comes from the minimal positive solution
that is found in the ranges $z^{*}\in(0,1/(1-q))$ when $|q|<1$ and
$z^{*}\in(0,\infty)$ when $q>1$. The integration contour can always
be deformed to pass through $z^{*}$ without crossing any singularities.
One can in principle transform the integration contour into the steepest
descent (stationary phase) contour, which explicit form however is
not easy to analyze. Instead, for the saddle point method to be applicable
it is enough to construct the steep descent contour, where the real
part of $h(z)$, 
\[
\Re h(z)=\left\{ \begin{array}{cc}
-\sum_{i=0}^{\infty}\ln|1-z(1-q)q^{i}|-\rho\ln|z|, & |q|<1;\\
\sum_{i=0}^{\infty}\ln|1-z(1-q)q^{-i-1}|-\rho\ln|z|, & |q|>1,
\end{array}\right.
\]
 monotonously decreases away from $z^{*}$. In the case $q>0$ the
simplest choice is the circle $|z|=z^{*},$which ensures the correct
sign of the derivative of the real part, 
\begin{eqnarray}
\frac{d}{d\phi}\Re h(z^{*}e^{i\phi}) & = & \left\{ \begin{array}{cc}
\sum_{i=0}^{\infty}\frac{-z^{*}sin\varphi(1-q)q^{i}}{|1-z^{*}e^{i\varphi}(1-q)q^{i}|^{2}}, & |q|<1;\\
\sum_{i=0}^{\infty}\frac{z^{*}sin\phi(1-q)q^{-i-1}}{|1-z^{*}e^{i\varphi}(1-q)q^{-i-1}|^{2}}, & |q|>1,
\end{array}\right.
\end{eqnarray}
with respect to the polar angle $\varphi\in(-\pi,\pi)$: 
\begin{eqnarray*}
\mathrm{sgn}\frac{d}{d\phi}\Re h(z^{*}e^{i\varphi}) & =- & \mathrm{sgn}\varphi.
\end{eqnarray*}
We do not have such an estimate for the negative values of $q$. However
we expect that the results of the saddle point approximation can be
analytically continued to this region. 

Deforming the integration contour to the steep descent contour we
apply the standard saddle point technique, which yields the desired
asymptotic approximation 
\begin{eqnarray}
\oint\frac{dz}{2\pi iz}e^{Nh(z)}g(z) & = & \frac{e^{Nh_{0}}}{\sqrt{2\pi Nh_{2}}}\label{eq:int}\\
 & \times & \left[g_{0}+\frac{1}{2N}\left(\frac{h_{3}g_{1}}{h_{2}^{2}}-\frac{g_{2}}{|h_{2}|}-\frac{5g_{0}h_{3}^{2}}{12|h_{2}|^{3}}+\frac{g_{0}h_{4}}{4h_{2}^{2}}\right)+O(N^{-2})\right].\nonumber 
\end{eqnarray}
of (\ref{def_I}) in terms $g_{k}=(z\partial_{z})^{k}g(z)|_{z=z^{*}}$
and $h_{k}=(z\partial_{z})^{k}h(z)|_{z=z^{*}}$. In particular, for
the asymptotics of the partition function given by (\ref{def_I})
with $g(z)=1$ we obtain (\ref{eq:Z asymp}). 

Most quantities we deal with, like the stationary state averages,
are given by the normalized contour integrals, those from (\ref{def_I})
multiplied by the inverse partition function. Below we write such
integrals as $\oint D_{N,p}(t)g(t)$, where for brevity we introduce
the following notation for \textit{normalized differential }
\[
D_{N,p}(t):=\frac{dt}{2\pi i}\frac{1}{Z(N,p)}\frac{F^{N}(t)}{t^{p+1}}.
\]
The expansion (\ref{eq:int}) of the original unnormalized integral
starts from the exponential in $N$ factor. The normalization makes
the generic normalized integral (with $N-$independent integrand $g(z)$)
of order of $O(1).$ The lower corrections are $O(N^{-k})$ with $k=1,2,\dots$.
In particular in the first two orders we have 
\begin{equation}
\oint D_{N,p}(t)g(t)=g_{0}+\frac{1}{2N}\left(\frac{h_{3}g_{1}}{h_{2}^{2}}-\frac{g_{2}}{|h_{2}|}\right)+O(N^{-2}).\label{eq:int D asymp}
\end{equation}
An example of the normalized integral is the mean particle current
given by 
\begin{equation}
j_{N}=\oint D_{N,p}(z)z.\label{j int D}
\end{equation}
Applying (\ref{eq:int D asymp}) we obtain formulas (\ref{eq:j_infty},\ref{eq: current fss}).
The other asymptotic approximations of the stationary state averages
listed in the subsection \ref{subsec:Stationary-state-and} are obtained
similarly. Our further aim is to obtain asymptotic estimates of the
diffusion coefficient.

\subsection{KPZ regime \ $|q|\protect\neq1$}

We first note that the formula (\ref{Answer}) can be rewritten in
the form 

\begin{eqnarray}
\Delta & = & N^{2}\frac{Z(2N,2p)}{Z(N,p)^{2}}\oint D_{2N,2p}(t)t\phi(t)\nonumber \\
 & + & N^{2}\varoiintop D_{N,p}(t)D_{N,p}(y)yt\left[\frac{\rho}{t}-\frac{\phi(t)-\phi(y)}{t-y}\right]\label{eq: Delta 1}\\
 & + & 2N^{2}\sum_{i=1}^{\infty}\varoiintop D_{N,p}(t)D_{N,p}(y)yt\frac{q^{\pm i}\phi(yq^{\pm i})+\phi(y)}{t-yq^{\pm i}}\nonumber 
\end{eqnarray}
(We remind that the plus and minus signs are taken for $|q|<1$ and
$q>1$.) The main advantage of this representation is the appearance
of the single integral in the first line. To motivate this we note
that the integrand of the first double integral in (\ref{Answer})
has a pole at $t=y$. Though it is not a problem for the integral
at finite $N$, since the integration contours are nested, it becomes
an obstacle for an application of the saddle point method, because
the variables take identical values at the critical point, 

\begin{equation}
y=t=z^{*}.\label{eq: critical point}
\end{equation}
 To overcome this difficulty we separate the pole contribution as
follows
\begin{eqnarray}
 & \varoiintop_{|y|<|t|} & D_{N,p}(t)D_{N,p}(y)\frac{yt\phi(y)}{t-y}\nonumber \\
 & = & \frac{1}{2}\Big(\varoiintop D_{N,p}(t)D_{N,p}(y)yt\frac{\phi(y)-\phi(t)}{t-y}+\frac{Z(2N,2p)}{Z(N,p)^{2}}\oint D_{2N,2p}(t)t\phi(t)\Big).\label{Pulling_contours}
\end{eqnarray}
To obtain this identity we first divide the integral in l.h.s. into
two identical parts and then exchange the integration variables, $t\leftrightarrow y$,
in one of the parts. In this way we obtain the sum of two double integrals,
where the contours of integration in the variables $y$ and $t$ are
circles nested in two opposite ways, $|y|<|t|$ in the first and $|t|<|y|$
in the second. Then, we deform the contours in the second integral
to restore back the nesting of the first one. On the way we pick up
the contribution from the pole $y=t$ in the form of the single integral,
which yields (\ref{Pulling_contours}). Note that the integrand of
the double integral in r.h.s. is now regular at $y=t$ and, hence,
the nesting of the contours is irrelevant. 

Returning to (\ref{eq: Delta 1}) we find the single integral of (\ref{Pulling_contours})
in the first line and the double integral in the second line of (\ref{eq: Delta 1}).
We also used the representation (\ref{j int D}) of $J$, where the
first term in the square brackets comes from. 

Now, we are going to analyze the r.h.s (\ref{eq: Delta 1}) evaluating
the integrals in the steepest descent approximation. In particular
we will do it term by term under the infinite sum in the last line.
This interchange of the limit $N\to\infty$ and the infinite summation
requires, however, a proper justification, which we return to in the
end. 

Let us first argue that the leading contribution to the diffusion
coefficient comes from the first line of (\ref{eq: Delta 1}), while
the second and third lines are of smaller order. First we observe
that the asymptotic expansions of the terms with double integrals
contain only integer powers of $N$, while the one of the term with
the single integral is in half-integer powers. Indeed, as we have
already noted, the normalized integrals, where the integrand (outside
of the normalized differential) is $N$-independent, is normally $O(1)$
containing only integer powers of $N$ in corrections. In our case
the function $\phi(z)$ does depend on $N$ via the current $J$,
which, however, being the normalized integral itself is also expanded
in integer powers on $N.$ As $\phi(z)$ depends linearly on $J$
its $N-$dependence does not spoil the power integrality or half-integrality
of our expansions. Taking into account that the ratio of partition
functions in the first line is $O\left(\sqrt{N}\right),$ we would
expect the first line to be $O\left(N^{5/2}\right)$ and the sum of
the second and third lines to be $O\left(N^{2}\right)$, both with
corrections obtained by decreasing the powers of $N$ by integers.
However, this is not the case, since in both cases the leading terms
vanish. According to (\ref{eq:int D asymp}), to obtain the leading
term of the asymptotic expansion of a normalized integral we just
substitute the critical values of arguments $y=t=z^{*}$ to the integrand
together with taking the limit $N\to\infty$. Thus, the leading term
of the first line vanishes, as 
\begin{equation}
\lim_{N\to\infty}\phi(z^{*})=0.\label{eq: phi(z^=00007B*=00007D)=00003D0}
\end{equation}
To show that the leading term of the sum of the second and third lines
vanish we first note that the second ratio in the second line turns
into the derivative when its two agruments coincide
\[
\left.\frac{\phi(y)-\phi(t)}{t-y}\right|_{t=y}=\phi'(y).
\]
Also we notice that the derivative of $\phi(z)$ can be expressed
in terms of the function itself via an interesting relation 

\begin{equation}
\phi'(z)=\frac{\rho}{j_{N}}(\phi(z)+1)^{2}-\frac{2}{z}\sum_{i=1}^{\infty}\frac{q^{i}\left(\phi(z)-\phi(zq^{i})\right)}{1-q^{i}},\label{eq:phi' relation}
\end{equation}
which is justified by a direct check. Evaluating the integrands at
$z^{*}$ with $\phi'(z^{*})$ reexpressed using (\ref{eq:phi' relation}),
taking the limit $N\to\infty$ and using (\ref{eq: phi(z^=00007B*=00007D)=00003D0},\ref{eq:j_infty})
we find that the $O(N^{2})$ asymptotics of the second plus the third
line in (\ref{eq: Delta 1}) vanishes. 

The next corrections to the two contributions are of order of $O(N^{3/2})$
and $O(N)$ respectively. Therefore it is enough to evaluate the first
order correction of the single integral term using the formulas (\ref{eq:Z asymp})
and (\ref{eq:int D asymp}), which yields 
\[
\Delta=\frac{N^{3/2}\sqrt{\pi}}{8\sqrt{h_{2}}}\left(\frac{\phi_{1}h_{3}}{|h_{2}|}-\phi_{2}\right)+O(N).
\]
This ensures (\ref{lambda_2_KPZ_regime-1}). 

To obtain another form of the same formula, we note that 
\begin{eqnarray*}
\oint D_{2N,2p}(t)t\phi(t) & = & \oint D_{2N,2p}(t)t\left(\frac{J}{p}(\ln F(t))'-1\right)\\
 & = & \frac{1}{Z(2N,2p)}\left(j_{N}Z(2N,2p)-Z(2N,2p-1)\right),
\end{eqnarray*}
where we applied the integration by parts to the first term in the
brackets under the integral. Substituting the result into the first
line of (\ref{eq: Delta 1}) we arrive at the alternative representation
(\ref{eq:Delta via fss}) of the main asymptotics of $\Delta$.

Finally let us justify the interchange of the limit $N\to\infty$
and the infinite sum in 
\begin{equation}
\sum_{i=1}^{\infty}\varoiintop D_{N,p}(t)D_{N,p}(y)yt\frac{q^{\pm i}\phi(yq^{\pm i})+\phi(y)}{t-yq^{\pm i}}.\label{eq:sum}
\end{equation}
 To this end, we expand the denominator of the ratio into the geometric
series and exchange the summation and integration,
\begin{eqnarray*}
 &  & \varoiintop D_{N,p}(t)D_{N,p}(y)yt\frac{q^{\pm i}\phi(yq^{\pm i})+\phi(y)}{t-yq^{\pm i}}\\
 &  & =\sum_{k=0}^{\infty}q^{\pm ik}\varoiintop D_{N,p}(t)D_{N,p}(y)y(q^{\pm i}\phi(yq^{\pm i})+\phi(y))\left(\frac{y}{t}\right)^{k}.
\end{eqnarray*}
Here we used the fact that the sum in $k$ is in fact finite, as at
most $p$ integrals in $y$ are nonzero due to the integrand having
a pole at the origin, while the others being analytic inside the integration
contour. Choosing the integration contours to be the circles $|z|=z^{*}$
centered at the origin, one can show that the double integral can
be bounded by an $N-$independent constant $C>0$ (We use the boundedness
of $\phi(z)$ in the disk $|z|<z^{*}$ and asymptotic equality of
steep descent integrals $\oint\left|D_{N,p}(t)\right|$ and $\oint D_{N,p}(t)$,
while for the latter $\lim_{N\to\infty}\oint D_{N,p}(t)=1$ due to
(\ref{eq:int D asymp})). Hence the summand in (\ref{eq:sum}) is
bounded by the summable function $Cq^{\pm i}/(1-q^{\pm i}),$ which
allows us to apply the dominated convergence theorem to justify the
change of the order of limit and summation. 

\subsection{KPZ - EW crossover}

Here we consider the asymptotic behaviour of the diffusion coefficient
in the scaling limit, where the limit $q\to1$ is taken together with
$N\to\infty$. 
\begin{equation}
N\rightarrow\infty,\quad p\rightarrow\infty,\quad\frac{p}{N}=\rho,\quad q=e^{-\frac{\alpha}{\sqrt{N}}}.\label{sc_asym_limit}
\end{equation}
We start with the following representation of the formula (\ref{Answer})
\begin{eqnarray}
\Delta & = & N^{2}\Biggl[\rho j_{N}+2\sum_{i=0}^{\infty}\varoiintop D_{N,p}(t)D_{N,p}(y)yt\frac{\phi(y)}{t-yq^{i}}\label{eq:Delta 2}\\
 &  & +2\sum_{i=1}^{\infty}\varoiintop D_{N,p}(t)D_{N,p}(y)yt\frac{q^{i}\phi(yq^{i})}{t-yq^{i}}\Biggr],\nonumber 
\end{eqnarray}
written for the case $|q|<1$. Below, for the sake of economy we consider
the case $|q|<1$, while the case $|q|>1$ is obtained by the change
$q\to1/q$ in the series coefficients. 

We want to transform this expression to the form of well convergent
series, which could be integrated integrated term by term. To this
end, we use the representation of the function $\phi(z)$,
\[
\phi(z)=\frac{J}{p}\sum_{i=0}^{\infty}\frac{\left(1-q\right)^{i+1}}{1-q^{i+1}}z^{i}-1,
\]
obtained from (\ref{def_phi}) by expanding the denominator as a geometric
series and then changing the summation order. Also we make the similar
expansion of the denominators under the integrals in (\ref{eq:Delta 2}).
Performing the summation in $i$ we obtain 
\begin{eqnarray*}
\Delta & = & N^{2}\rho j_{N}+2N^{2}\frac{j_{N}}{\rho}\Biggl[\sum_{l=2}^{\infty}\frac{\left(1-q\right)^{l}}{1-q^{l}}\sum_{k=1}^{l-1}\frac{1}{1-q^{k}}\varoiintop D_{N,p}(t)D_{N,p}(y)t^{-k}y^{k+l}\\
 & + & \sum_{l=2}^{\infty}\frac{\left(1-q\right)^{l}}{1-q^{l}}\sum_{k=l}^{\infty}\frac{1}{1-q^{k}}\varoiintop D_{N,p}(t)D_{N,p}(y)\left(\frac{y}{t}\right)^{k}\left(y^{l}+t^{l}q^{k}\right)\Biggr]\\
 & + & 2N^{2}\left(\frac{j_{N}}{\rho}-1\right)\sum_{k=1}^{\infty}\frac{1}{1-q^{k}}\varoiintop D_{N,p}(t)D_{N,p}(y)\left(\frac{y}{t}\right)^{k}\left(y+tq^{k}\right),
\end{eqnarray*}
where we separated the sum in the first line from that in the second
line, to collect the $l-$th powers of $y$ and $t$ together. Remembering
that $(1-q)=O\left(1/\sqrt{N}\right)$ we notice that the external
summations in $l$ in the two first lines are already asymptotic in
nature due to the coefficients $\left(1-q\right)^{l}/\left(1-q^{l}\right)=O\left(N^{-\left(l-1\right)/2}\right)$,
each term being $O\left(N^{-1/2}\right)$ times the previous one.
On the other hand, we will see that there are $O\left(\sqrt{N}\right)$
terms of the same order inside the sums in $k$, which bring the dominant
contribution to these sums. Therefore, it is enough to limit our consideration
by a few first values of $l$. Specifically, we leave the terms with
$l=2,3,4$ in the first line and the terms with $l=2,\dots,5$ in
the second line to find the expansion up to the order $O(N)$. Then,
only finitely many terms remains in the first line and five infinite
sums in the second and the third lines. We also separate finitely
many terms from the latters, to unify the infinite sums into a single
infinite sum starting from the index value $k=5$. Thus, asymptotically
up to the terms of order of $O(N)$ the diffusion coefficient can
be represented as consisting of two parts, 

\begin{equation}
\Delta=FS+IS+O\left(\sqrt{N}\right),\label{eq:Delta=00003DFS+IS}
\end{equation}
a finite sum 
\begin{eqnarray}
FS & = & N^{2}\Biggl[\rho j_{N}+2\frac{j_{N}}{\rho}\sum_{l=2}^{4}\frac{\left(1-q\right)^{l}}{1-q^{l}}\left(\sum_{k=1}^{l-1}\frac{J_{k}^{l}}{1-q^{k}}+\sum_{k=l}^{4}\frac{I_{k}^{l}}{1-q^{k}}\right)\label{eq:FS}\\
 &  & +2\left(\frac{J}{p}-1\right)\sum_{k=1}^{4}\frac{I_{k}^{1}}{1-q^{k}}\Biggr]\nonumber 
\end{eqnarray}
and an infinite sum
\begin{equation}
IS=2N^{2}\sum_{k=5}^{\infty}\frac{1}{1-q^{k}}\left[\left(\frac{j_{N}}{\rho}-1\right)I_{k}^{1}+\frac{j_{N}}{\rho}\left(\frac{\left(1-q\right)^{2}}{1-q^{2}}I_{k}^{2}+\cdots+\frac{\left(1-q\right)^{5}}{1-q^{5}}I_{k}^{5}\right)\right],\label{eq:IS}
\end{equation}
where for brevity we introduce the following notations for two products
of contour integrals appearing repeatedly in the calculations

\begin{eqnarray}
J_{k}^{l} & = & \varoiintop D_{N,p}(t)D_{N,p}(y)\left(\frac{y}{t}\right)^{k}y^{l},\label{eq: J_k^L}\\
I_{k}^{l} & = & \varoiintop D_{N,p}(t)D_{N,p}(y)\left(\frac{y}{t}\right)^{k}\left(y^{l}+t^{l}q^{k}\right)=J_{k}^{l}+q^{k}J_{k-l}^{l}.\label{eq:I_k^l}
\end{eqnarray}

The finite sum is easy to calculate asymptotically using the formula
(\ref{eq:int D asymp}) for the normalized integrals, which yields
\begin{eqnarray}
J_{k}^{l} & = & \left(z^{*}\right)^{l}\left(1+\frac{1}{2N}\left(\frac{h_{3}l}{h_{2}^{2}}-\frac{\left(k+l\right)^{2}+k^{2}}{|h_{2}|}\right)\right).\label{eq:J_k^l asymp}
\end{eqnarray}
The ingredients of this formula are expressed in terms of the saddle
point $z^{*}$, which can asymptotically be found in the limit (\ref{sc_asym_limit})
from perturbative solution of the saddle point equation as the expansion
in powers of $N^{-1/2}$. The solution in the four first orders yields
\begin{equation}
z^{*}=\rho-\frac{\alpha\rho^{2}}{2\sqrt{N}}+\frac{\alpha^{2}\rho^{3}}{6N}-\frac{\alpha^{3}\rho^{2}\left(\rho^{2}-1\right)}{24N^{3/2}}+O\left(N^{-2}\right).\label{eq:z^* asymp}
\end{equation}
Then, the expansion of the derivatives of function $h(z)$ follows
\begin{eqnarray}
h_{2}=\rho+\frac{\alpha\rho^{2}}{2\sqrt{N}}+\frac{\alpha^{2}\rho^{3}}{6N}+\frac{-\alpha^{3}\rho^{2}+\alpha^{3}\rho^{4}}{24N^{3/2}}+O(N^{-2}),\label{eq: h_2 asymp}\\
h_{3}=\rho+\frac{3\alpha\rho^{2}}{2\sqrt{N}}+\frac{7\alpha^{2}\rho^{3}}{6N}+\frac{-\alpha^{3}\rho^{2}+5\alpha^{3}\rho^{4}}{8N^{3/2}}+O(N^{-2}).\label{eq:h_3 asymp}
\end{eqnarray}
The expansions of $J_{k}^{l}$ and hence $I_{k}^{l}$ for finite values
of $k$ and $l$ appearing in (\ref{eq:FS}) can be obtained by substitution
of (\ref{eq:z^* asymp}-\ref{eq:h_3 asymp}) into (\ref{eq:J_k^l asymp}).
In particular for of the particle current we have 
\begin{eqnarray}
j_{N} & = & J_{0}^{1}=z^{*}\left(1+\frac{1}{2N}\left(\frac{h_{3}}{h_{2}^{2}}-\frac{1}{|h_{2}|}\right)\right)\label{as_J}\\
 & = & \rho-\frac{\alpha\rho^{2}}{2\sqrt{N}}+\frac{\alpha^{2}\rho^{3}}{6N}+\frac{1}{N^{3/2}}\left(\frac{\alpha\rho}{2}-\frac{1}{24}\alpha^{3}\rho^{2}\left(\rho^{2}-1\right)\right)+O\left(N^{-2}\right).
\end{eqnarray}
Substituting them into (\ref{eq:FS}) we obtain the approximation
of the finite sum

\begin{equation}
FS=\frac{\alpha\rho^{2}}{2}N^{3/2}-\frac{\alpha^{2}\rho^{3}}{12}N+O\left(\sqrt{N}\right),\label{eq:FS asymp}
\end{equation}
where the terms of order of $O(N^{2})$ have canceled. 

It is more tricky to analyze the infinite sum. First we note that
under a closer look one finds that the infinite sum is in fact finite,
since $I_{k}^{l}$ vanishes when $k>p+l-1$. However, the summation
index $k$ becomes unbounded in the limit $p\to\infty,$ and its value
affects the location of the saddle points in the integrals. To account
for this effect let us write $I_{k}^{l}$ in the form.
\begin{equation}
I_{k}^{l}=\frac{Z(N,p-k)Z(N,p+k)}{Z(N,p)^{2}}\varoiintop D_{N,p+k}(t)D_{N,p-k}(y)\left(y^{l}+t^{l}q^{k}\right).\label{eq:I_k^l shifted}
\end{equation}
We now claim that the coefficient before the integral is responsible
for the effective range of the index $k$, where the main contribution
to the infinite sum (\ref{eq:IS}) comes from. To this end, we introduce
the free energy 
\[
\mathfrak{f}(\rho)=-\lim_{N\to\infty}N^{-1}\ln Z(N,[\rho N])=-h(z^{*}).
\]
 Then the ratio of partition functions of interest is approximated
by 
\begin{eqnarray}
\frac{Z(N,p-k)Z(N,p+k)}{Z(N,p)^{2}} & \simeq & \exp\left[-N\left(\mathfrak{f}\left(\rho+\frac{k}{N}\right)+\mathfrak{f}\left(\rho-\frac{k}{N}\right)-2\mathfrak{f}(\rho)\right)\right]\label{eq:Z-ratio}\\
 & \simeq & \exp\left(-\frac{\mathfrak{f}''(\rho)k^{2}}{N}\right)=\exp\left(-\frac{k^{2}}{Nh_{2}}\right).\nonumber 
\end{eqnarray}
where we used the fact that 
\begin{eqnarray*}
\mathfrak{f}''(\rho) & =- & \frac{d}{d\rho}\left(\frac{dz^{*}}{d\rho}\frac{\partial h(z^{*})}{\partial z^{*}}+\frac{\partial h(z^{*})}{\partial\rho}\right)=\frac{dz^{*}}{d\rho}\frac{\partial\ln z^{*}}{\partial z^{*}}=\frac{1}{h_{2}}
\end{eqnarray*}

The estimate (\ref{eq:Z-ratio}) suggests that the main contribution
to the infinite sum (\ref{eq:IS}) comes from the scale, in which
$k$ is of order of $\sqrt{N}$. It is not difficult also to find
corrections to the exponent. There are two sources of corrections.
One source is from approximation of the $\ln Z(N,[\rho N])$ by the
free energy. The other is from approximation of the expressions at
shifted densities $\rho\pm k/N$ with those at density $\rho$. Their
orders of magnitude are $O\left(k^{2}N^{-2}\right)$ and $O\left(k^{4}N^{-3}\right)$
respectively. Note that due to the symmetry of the exact expression
with respect to the transformation $k\leftrightarrow-k$ the corrections
contain only even powers of $k$. If we also replace $h_{2}$ in the
denominator of the exponent by its limiting value $\rho,$ we also
acquire the correction of order of $\left(k^{2}N^{-3/2}\right)$,
which is larger of the two former. Since the latter correction is
relevant for our further calculations, we leave its explicit value
extracted from (\ref{eq: h_2 asymp}) in the final expression, which
therefore looks as follows 
\[
\frac{Z(N,p-k)Z(N,p+k)}{Z(N,p)^{2}}=\exp\left(-\frac{k^{2}}{N\rho}-\frac{\alpha k^{2}}{2N^{3/2}}+O\left(\frac{k^{2}}{N^{2}}\right)+O\left(\frac{k^{4}}{N^{3}}\right)\right)
\]
 Apparently this approximation works well for not too large values
of $k$. We will use it in the range $0<k<\epsilon N$, where $\epsilon$
is a constant that can be chosen arbitrarily small. 

For $k>N\epsilon$ the ratio is at least exponentially small in $N$.
This follows from the monotonicity of the exponent in (\ref{eq:IS})
as the function of the ratio $(k/N)$. Indeed for $0<x<\rho$ we have
\begin{eqnarray*}
\frac{d}{dx}\left(\mathfrak{f}\left(\rho+x\right)+\mathfrak{f}\left(\rho-x\right)-2\mathfrak{f(}\rho))\right) & = & \left.\frac{\partial h(z^{*})}{\partial\rho}\right|_{\rho\to\rho+x}-\left.\frac{\partial h(z^{*})}{\partial\rho}\right|_{\rho\to\rho-x}\\
 &  & \!\!\!\!\!\!\!\!\!\!\!\!\!\!\!\!\!\!\!\!\!\!\!\!\!\!\!\!\!\!\!\!\!\!\!\!\!\!\!\!\!\!\!\!\!\!\!\!\!\!\!\!\!\!\!\!\!\!\!\!\!\!\!\!=\left.\ln z^{*}\right|_{\rho\to\rho+x}-\left.\ln z^{*}\right|_{\rho\to\rho+x}=2x\left.\frac{d\ln z^{*}}{d\rho}\right|_{\rho\to\rho'}=2x\left.\frac{z^{*}}{h_{2}}\right|_{\rho\to\rho'}>0,
\end{eqnarray*}
where the subscripts show that the corresponding expressions are evaluated
at shifted values of the density. We also used the mean value theorem,
to approximate the function $\ln z^{*}$ of the density in the interval
$[\rho-x,\rho+x]$ by the linear function with the slope equal to
the derivative of $\ln z^{*}$ in $\rho$ at some intermediate point
$\rho'\in[\rho-x,\rho+x]$. From the monotonicity of the exponent
we conclude that for any small $\epsilon>0$ there exists a constant
$c$, such that 
\begin{equation}
\frac{Z(N,p-k)Z(N,p+k)}{Z(N,p)^{2}}=O\left(e^{-cN}\right),\quad k>\epsilon N.\label{eq:k>epsilon N}
\end{equation}

It remains to evaluate the double integral in (\ref{eq:I_k^l shifted})
asymptotically. To this end, we must take into account that the dominating
critical points in the integrals in $y$ and $t$ are now shifted
due to the density shifts $\rho\to\rho\pm k/N.$ 
\begin{eqnarray*}
\varoiintop D_{N,p+k}(t)D_{N,p-k}(y)\left(y^{l}+t^{l}q^{k}\right) & = & \left[J_{0}^{l}\right]_{\rho\to\left(\rho-k/N\right)}+q^{k}\left[J_{0}^{l}\right]_{\rho\to\left(\rho+k/N\right)}
\end{eqnarray*}
Then, we use the asymptotic formulas (\ref{eq:J_k^l asymp})-(\ref{eq:h_3 asymp})
for $J_{k}^{l}$ evaluated at shifted densities. The expression obtained
is rather cumbersome and we omit its explicit form here. Instead,
we show the asymptotic form of the summand of (\ref{eq:IS}), which
simplifies significantly due to many cancellations,

\begin{eqnarray}
[IS]_{k} & = & 2N^{2}\exp\left(-\frac{k^{2}}{N\rho}-\frac{\alpha k^{2}}{2N^{3/2}}+O\left(\frac{k^{2}}{N^{2}}\right)+O\left(\frac{k^{4}}{N^{3}}\right)\right)\nonumber \\
 & \times & \Biggl[-\frac{\alpha\rho}{2}\left(\frac{k}{\sqrt{N}}\right)\frac{1}{N}+\left(\frac{\alpha^{2}\rho^{2}}{3}\left(\frac{k}{\sqrt{N}}\right)+\frac{\alpha}{2}\frac{\left(1+q^{k}\right)}{\left(1-q^{k}\right)}\left(\frac{k}{\sqrt{N}}\right)^{2}\right)\frac{1}{N^{3/2}}\label{eq:=00005BIS=00005D_k}\\
 &  & +O\left(\frac{1}{N^{2}}\right)+O\left(\frac{k}{N^{5/2}}\right)+O\left(\frac{k^{2}}{N^{3}}\right)+O\left(\frac{k^{3}}{N^{7/2}}\right)\Biggr].\nonumber 
\end{eqnarray}
Here we implied that $k$ is of order of $\sqrt{N}$, and thus the
exact value of $q^{k}$ is kept. 

Now we are in a position to evaluate the infinite sum asymptotically.
To this end, we divide it into two parts 
\begin{eqnarray*}
IS & = & \sum_{k=5}^{[\epsilon N]}[IS]_{k}+O(e^{-c_{1}N}).
\end{eqnarray*}
The second part is exponentially small in $N$ with some constant
$c_{1}>0$ due to (\ref{eq:k>epsilon N}) and the fact that there
is at most $p$ nonzero summands in the whole sum. The first sum can
be approximated by an integral 
\begin{eqnarray}
 & \sum_{k=5}^{[\epsilon N]}[IS]_{k} & =2N^{2}\sqrt{N\rho}\int_{0}^{\infty}e^{-x^{2}}\Biggl[-\frac{\alpha x\rho^{3/2}}{2}\frac{1}{N}\label{eq: sum =00005BIs=00005D_k}\\
 & + & \!\!\!\!\!\!\!\!\!\!\!\!\!\!\!\!\left(\frac{\alpha^{2}\rho^{5/2}}{3}x+\frac{\alpha\rho}{2}\frac{1+e^{-\alpha x\sqrt{\rho}}}{1-e^{-\alpha x\sqrt{\rho}}}x^{2}-x^{3}\frac{\alpha^{2}\rho^{5/2}}{4}\right)\frac{1}{N^{3/2}}\Biggr]dx+O\left(\sqrt{N}\right)\nonumber \\
 & = & \!\!\!\!\!\!\!\!\!\!\!\!\!\!\!\!\!-\frac{\alpha\rho^{2}N^{3/2}}{2}+\left(\frac{\alpha^{2}\rho^{3}}{12}+\alpha\rho^{3/2}\int_{0}^{\infty}e^{-x^{2}}\frac{1+e^{-\alpha x\sqrt{\rho}}}{1-e^{-\alpha x\sqrt{\rho}}}x^{2}dx\right)N+O\left(\sqrt{N}\right).\nonumber 
\end{eqnarray}
Going from (\ref{eq:=00005BIS=00005D_k}) to (\ref{eq: sum =00005BIs=00005D_k})
we left only the first term in the exponential factor. The exponential
of the other three terms are replaced by their expansion to the second
order. As a result we obtain an extra term comparing to the sum in
the square brackets in (\ref{eq: sum =00005BIs=00005D_k}) and several
more correction terms under the sum. Finally, we replace the variables
in the integral to $x=k^{2}/(N\rho)$. All the integrals except one
can be evaluated explicitly, while one remains in the integral form. 

To estimate the overall corrections we need to evaluate the difference
of the exact expression (with corrections) and the limiting expression
(without corrections) and make sure that the resulting sum, which
we extend back to infinity converges uniformly in $N$, so that the
summation and the limit $N\to\infty$ can be interchanged. The uniform
convergence is ensured by the Gaussian prefactor, whose exponent with
corrections can always be kept negative by the choice of small enough
$\epsilon.$ The explicit calculation is standard, and we omit the
details. In addition one has to take into account the Euler-Maclaurin
corrections coming from the integral approximation of the sum. The
resulting corrections have an order $O\left(\sqrt{N}\right).$ 

Finally, when we add the finite and infinite sums all the terms except
one integral surprisingly cancel, yielding 
\[
\Delta=N\alpha\rho^{3/2}\int_{0}^{\infty}e^{-x^{2}}\frac{1+e^{-\alpha x\sqrt{\rho}}}{1-e^{-\alpha x\sqrt{\rho}}}x^{2}dx+O\left(\sqrt{N}\right),
\]
which agrees with (\ref{eq:F(g,infty)},\ref{Sc_limit<1}). Following
the same way in the case $|q|>1$ we observe that the change $q\leftrightarrow1/q$
corresponds to the change $\alpha\leftrightarrow-\alpha$ in the final
expression, which leaves it invariant. Thus, the formula obtained
is valid for approaching the limit $q\to1$ from both sides.

\section{Conclusion.}

To summarize, we have obtained the exact integral representation for
the diffusion constant of the q-boson ZRP. Asymptotic analysis of
the result obtained yielded the expressions of the diffusion coefficient
in the KPZ regime as well as KPZ-EW crossover scaling function. Both
asymptotic results agree with the scaling hypothesis about KPZ universality
and earlier solutions of other models. Our results in principle should
follow from the recent finite-time results on q-boson ZRP \citep{LSW}.
However, the connection is highly nontrivial.  

This work can be considered a first step in the studies of the large
time behaviour of q-boson type models in confined geometry. The next
goal could be a construction of the higher cumulants of the model.
In particular, it would be interesting to study KPZ-EW crossover not
only for particular current cumulants, but also for the whole large
deviation function.

Also richer structure of the q-boson-like models with more parameters,
e.g. of the chipping model with factorized steady state also called
q-Hahn or $q,\mu,\nu$-ZRP, opens perspectives for studies of transitions
between KPZ and other than EW types of behaviour. We mention the examples
of two particular limits of that model with such a transitions. First,
the jamming transition in the TASEP with generalized update was recently
studied in  \cite{DPP2015}. This model is of the same type as TASEP corresponding
to the quantum group parameter $q=0$. The Bethe ansatz solution simplifies
greatly in that case, allowing obtaining the whole scaled cumulant
generating function by the Derrida-Lebowitz method. It crosses over
from the KPZ to Gaussian deterministic aggregation regime. We expect
that similar transition must also take place in the chipping model
\citep{P2013} with $q\neq0.$ In the latter case however the method
of Derrida-Lebowitz is inapplicable. The appropriate tool would then
be the T-Q relation approach developed in \citep{PM2008} adopted
for the q-boson models in the present paper. 

Also, the asymmetric avalanche process \cite{PIPH2001,PPH2003} gives an example of the transition
from dispersive to continuous avalanche flow. The latter model is
the stochastic particle model with particle flow maintained by the
non-local avalanche events. The avalanches are finite in the infinite
system at low particle densities, while the mean avalanche size diverges
as the density approaches the critical value. In the context of dynamics
of the associated interface the system suffers the transition from
the KPZ to the tilted interface universality class \cite{TKD,PPH2003_2}. How the current
diffusion coefficient describes the crossover between these two regimes
is still an open problem.

Finally, we mention the similar problem for q-boson ZRP on the open
segment. While the current-density diagram was obtained for this model
using the matrix product representation of the stationary state, neither
the further current cumulants nor the large deviation function was yet obtained. Development of the T-Q approach to this problem
is a challenging problem. Similarly to the case of periodic system we expect that in the universal scaling limits the results will match with those obtained from the solution of ASEP with open boundaries \citep{DEM2,LMV2012}. This is a matter of further studies.

\ack{}{}

The work is supported by Russian Foundation from Basic Research under
grant 19-01-00726 A. For
the first author, the study has been funded by the Russian Academic Excellence Project '5-100'. We are indebted to Maria Matushko, who participated
in the early stages of this work. We are grateful to Pavel Pyatov
and Sergey Khoroshkin for fruitful discussions.

\appendix

\section{Derivation of equation (\ref{lambda2 before integration on z}) \label{sec:Derivation-of-equation}}

We simplify the result (\ref{Prelim_lambda2}) by substituting 
\begin{equation}
f(x)=NB_{1}(qx)+\frac{Q_{1}(x)}{x^{p}}[(1-x)^{N}B_{1}(x)-B_{1}(qx)]\label{fviaBQ}
\end{equation}
to the integral 
\[
\sum_{i=0}^{\infty}\oint\frac{F^{N}(z)}{z^{p}}\phi(z)\frac{f(q^{i}(1-q)z)}{F(q^{i+1}z)^{N}}\frac{dz}{2\pi\mathrm{i}}.
\]
The first summand of $f(x)$ does not contribute to the $\lambda_{2}$.
Indeed, for any $i$ we have 
\begin{eqnarray*}
 & \oint & \frac{F(z)^{N}}{z^{p}}\phi(z)\frac{NB_{1}(q^{i+1}(1-q)z)}{F(q^{i+1}z)^{N}}\frac{dz}{2\pi\mathrm{i}}\\
 &  & =-\frac{N(1-q)^{p}}{Z(N,p)}\oint\frac{NF(z)^{N}}{z^{p}}\Big(\frac{N}{p}\frac{Z(N,p-1)}{Z(N,p)}\frac{F'(z)}{F(z)}-1\Big)\frac{dz}{2\pi\mathrm{i}}=0
\end{eqnarray*}
where we first replace the degree $(p-1)$ polynomial $B_{1}(q^{i+1}(1-q)z)$,
which is the truncation of the function $N(1-q)^{p}F(q^{i+1}z)^{N}/Z(N,p),$
by the whole function that cancels the denominator. Apparently, this
replacement does not affect the value of the integral. Indeed, integrating
the series representation of the function term by term we find that
the integrals with the terms of higher than $(p-1)$ order vanish
for the integrands being analytic inside the integration contour.
Substituting the explicit form of $\phi(z)$ and integrating by parts
the first term in the brackets in the second line we see that the
whole integral vanishes. 

Substitution of the second part of $f(x)$ into the integral under
the sum in (\ref{Prelim_lambda2}) gives 
\begin{eqnarray}
\frac{1}{(1-q)^{p}}\oint\frac{dz}{2\pi i}\frac{\phi(z)}{z^{2p}}\sum_{i=0}^{\infty}\frac{Q_{1}((1-q)q^{i}z)}{(1-q)^{p}q^{ip}}(F_{0i}(z)B_{1}((1-q)q^{i}z)\label{E2viaB1}\\
-F_{0i+1}(z)B_{1}((1-q)q^{i+1}z))=\nonumber 
\end{eqnarray}
\begin{eqnarray*}
=\frac{1}{(1-q)^{2p}}\oint\frac{dz}{2\pi i}\frac{\phi(z)}{z^{2p}}\sum_{i=0}^{\infty}\frac{Q_{1}((1-q)q^{i}z)}{q^{ip}}\left[F_{0i+1}(z)B_{1}^{\geq p}((1-q)q^{i+1}z)\right.\\
\left.-F_{0i}(z)B_{1}^{\geq p}((1-q)q^{i}z))\right].
\end{eqnarray*}
where we introduce
\begin{eqnarray}
B_{1}^{\geq p}((1-q)x) & = & -\frac{N(1-q)^{p}}{Z(N,p)}F^{N}(x)-B_{1}((1-q)x)\label{B_1tail}\\
 & = & -\frac{N(1-q)^{p}}{Z(N,p)}x^{p}\oint_{|x|<|z|}\frac{dz}{2\pi i}\frac{F^{N}(z)}{z^{p}(z-x)}.\nonumber 
\end{eqnarray}
The expression the square brackets in (\ref{E2viaB1}) is the sum
of two functions with successive indices. Collecting the terms with
the same index from two successive summands of the sum in index $i$
we obtain 
\begin{eqnarray*}
 & = & -\frac{1}{(1-q)^{2p}}\oint\frac{dz}{2\pi i}\frac{\phi(z)}{z^{2p}}Q_{1}((1-q)z)B_{1}^{\geq p}((1-q)z)\\
 & + & \frac{1}{(1-q)^{2p}}\oint\frac{dz}{2\pi i}\frac{\phi(z)}{z^{2p}}\sum_{i=0}^{\infty}B_{1}^{\geq p}((1-q)q^{i+1}z)F_{0i+1}(z)\\
 & \times & \left[\frac{Q_{1}((1-q)q^{i}z)}{q^{ip}}-\frac{Q_{1}((1-q)q^{i+1}z)}{q^{(i+1)p}}\right].
\end{eqnarray*}
The content of the square brackets in the latter formula can be further
modified to 
\begin{eqnarray*}
 &  & \frac{Q_{1}((1-q)q^{i}z)}{q^{ip}}-\frac{Q_{1}((1-q)q^{i+1}z)}{q^{(i+1)p}}\\
 &  & =\frac{N(1-q)^{p}q^{-i}z^{p}}{Z(N,p)}\oint\frac{F^{N}(y)}{y^{p}}\frac{dy}{2\pi i}(\sum_{k=0}^{\infty}\frac{q^{k}}{z-yq^{k-i}}-\sum_{k=0}^{\infty}\frac{q^{k-1}}{z-yq^{k-i-1}})=\\
 &  & =-\frac{N(1-q)^{p}q^{-i}z^{p}}{Z(N,p)}\oint\limits _{|y|<|q^{i+1}z|}\frac{F^{N}(y)}{y^{p}}\frac{dy}{2\pi i}\frac{q^{-1}}{z-yq^{-i-1}},
\end{eqnarray*}
where we used the integral form (\ref{Q1(x) int form}) of $Q_{1}(x)$.
Substituting also the integral of $B_{1}^{\geq p}((1-q)x)$ from (\eqref{B_1tail})
and inserting the resulting integral into (\ref{Prelim_lambda2})
we arrive at (\ref{lambda2 before integration on z}).

\end{document}